\documentclass[aps,prd,10pt,reprint,superscriptaddress,preprintnumbers,nofootinbib,floatfix,longbibliography]{revtex4-1}

\usepackage[T1]{fontenc}
\usepackage[utf8]{inputenc}
\usepackage[english]{babel}
\usepackage{amsmath,amsthm,amssymb,amsfonts,mathrsfs,amsbsy,bm}
\usepackage{comment}
\usepackage{tensor}
\usepackage{slashed}
\usepackage{float}
 \usepackage{mathrsfs,tikz} 
\usepackage{esint}
\usepackage[a4paper, margin=1.3cm]{geometry}
\usepackage{cancel}
\usepackage{graphicx}
\usepackage{multirow}
\usepackage{array}
\usepackage{booktabs}
\usepackage{makecell}
\usepackage{xcolor}
\colorlet{BLUE}{blue}
\usepackage{tikz}
\usetikzlibrary{quotes,angles,arrows,decorations.markings,decorations.pathmorphing}
\usepackage{hyperref}
\hypersetup{colorlinks=true,breaklinks=true,citecolor=blue,linkcolor=[rgb]{0,0.5,0.9},urlcolor=blue}
\graphicspath{{figures/}}

\newcommand{\be}{\begin{equation}}
\newcommand{\ee}{\end{equation}}
\newcommand{\Be}{\begin{eqnarray}}
\newcommand{\Ee}{\end{eqnarray}}

\newcommand{\mincir}{\raise-3.truept\hbox{\rlap{\hbox{$\sim$}}\raise4.truept\hbox{$<$}\ }}
\newcommand{\magcir}{\raise-3.truept\hbox{\rlap{\hbox{$\sim$}}\raise4.truept\hbox{$>$}\ }}

\providecommand{\U}[1]{}

\newcommand{\ie}{\begin{equation}}
\newcommand{\fe}{\end{equation}}
\newcommand{\se}{\begin{eqnarray}}
\newcommand{\ff}{\end{eqnarray}}

\begin{document}
\emergencystretch=4em
\usetikzlibrary{arrows.meta,patterns}

%%%%%%%%%%%%%%%%%%%%%%%%%%%%%%%%%%%%%%%%%%%%%%%%%%%%%%%%%%%%%%%%%%%%%%%%%%%%%%%%%%%%%%%%%%%%%%%%%%%%%%%%%%%%%%%%%%%%%%%%%%%%%%%%%%%%%%%%%%%%%%%%%%%%%%%%%%%%%%%%%%%%%%%%%%%%%%%%%%%%%%%%%%%%%%%%%%%%%%%%%%%%%%%%%%%%%%%%%%%%%%%%%%%%%%%%%%%%%%%%%%%%%%%%%%%%%%%%%%%%%%%%%%%%%%%%%%%%%%%%%%%%%%%%%%%%%%%%%%%%%%%%%%%%%%%%%%%%%%%%%%%%%%%%%%%%%%%%%%%%%%%%%%%%%%%%%%%%%%%%%%%%%%%%%%%%%%%%%%%%%%%%%%%%%%%%%%%%%%%%%%%%%%%%%%%%%%%%%%%%%%%%%%%%%%%%%%%%%%%%%%%%%%%%%%%%%%%%%%%%

\title{Bumblebee radiation from neutral matter around a black hole}

\author{A. A. Ara\'{u}jo Filho}
\email{dilto@fisica.ufc.br}
\affiliation{Departamento de F\'isica, Universidade Federal da Para\'iba, Caixa Postal 5008, 58051--970, Jo\~ao Pessoa, Para\'iba, Brazil.}
\affiliation{Departamento de F\'isica, Universidade Federal de Campina Grande, Caixa Postal 10071, 58429--900 Campina Grande, Para\'iba, Brazil.}
\affiliation{Center for Theoretical Physics, Khazar University, 41 Mehseti Street, Baku, AZ-1096, Azerbaijan.}

%%%%%%%%%%%%%%%%%%%%%%%%%%%%%%%%%%%%%%%%%%%%%%%%%%%%%%%%%%%%%%%%%%%%%%%%%%%%%%%%%%%%%%%%%%%%%%%%%%%%%%%%%%%%%%%%%%%%%%%%%%%%%%%%%%%%%%%%%%%%%%%%%%%%%%%%%%%%%%%%%%%%%%%%%%%%%%%%%%%%%%%%%%%%%%%%%%%%%%%%%%%%%%%%%%%%%%%%%%%%%%%%%%%%%%%%%%%%%%%%%%%%%%%%%%%%%%%%%%%%%%%%%%%%%%%%%%%%%%%%%%%%%%%%%%%%%%%%%%%%%%%%%%%%%%%%%%%%%%%%%%%%%%%%%%%%%%%%%%%%%%%%%%%%%%%%%%%%%%%%

\date{\today}

%%%%%%%%%%%%%%%%%%%%%%%%%%%%%%%%%%%%%%%%%%%%%%%%%%%%%%%%%%%%%%%%%%%%%%%%%%%%%%%%%%%%%%%%%%%%%%%%%%%%%%%%%%%%%%%%%%%%%%%%%%%%%%%%%%%%%%%%%%%%%%%%%%%%%%%%%%%%%%%%%%%%%%%%%%%%%%%%%%%%%%%%%%%%%%%%%%%%%%%%%%%%%%%%%%%%%%%%%%%%%%%%%%%%%%%%%%%%%%%%%%%%%%%%%%%%%%%%%%%%%%%%%%%%%%%%%%%%%%%%%%%%%%%%%%%%%%%%%%%%%%%%%%%%%%%%%%%%%%%%%%%%%%%%%%%%%%%%%%%%%%%%%%%%%%%%%%%%%%%%%%%%%%%%%%%%%%%%%%%%%%%%%%%%%%%%%%%%%%%%%%%%%%%%%%%%%%%%%%%%%%%%%%%%%%%%%%%%%%%%%%%%%%%%%%%%%%%%%%%%%%%%%%%%%%%%%%%%%%%%%%%%%%%%%%%%%%%%%%%%%%%%%%

\begin{abstract}

We investigate bumblebee radiation generated by neutral matter orbiting a black hole with a radial spacelike vector vacuum. Starting from a point particle coupled minimally to the metric, we derive the curvature induced vector current and the potential response required by its conservation. For a localized circular source with no independent radial boundary current, this response gives $\delta V'=\xi\rho/4$. The axial vector source instead follows the radial transport of angular momentum and vanishes for circular motion. A slightly eccentric orbit excites an axial dipole at the orbital frequency $\Omega$ and a quadrupole contribution at $2\Omega$, with outgoing amplitudes proportional to $ev^3$ and $ev^4$, respectively, where $v=\sqrt{M/a}$. We construct their retarded response and obtain the leading coefficients through weak field matching. Reconstruction of the observer's tidal tensor gives the dipole strain $h_{\hat r\hat\phi}^{V,10}=\xi^2b^2\mu ev^3\sin\theta\cos[\Omega(t-R)]/(8\pi R)$. The corresponding partial luminosity is $P_{B,10}^{\infty}=\xi^2b^2\mu^2e^2v^{12}/(48\pi M^2)$ at the leading order. Remarkably, the ordinary neutral matter produces an additional radial transverse tidal signal even when the homogeneous gravitational master operator remains in Schwarzschild form.

\end{abstract}

\maketitle

\tableofcontents

%%%%%%%%%%%%%%%%%%%%%%%%%%%%%%%%%%%%%%%%%%%%%%%%%%%%%%%%%%%%%%%%%%%%%%%%%%%%%%%%%%%%%%%%%%%%%%%%%%%%%%%%%%%%%%%%%%%%%%%%%%%%%%%%%%%%%%%%%%%%%%%%%%%%%%%%%%%%%%%%%%%%%%%%%%%%%%%%%%%%%%%%%%%%%%%%%%%%%%%%%%%%%%%%%%%%%%%%%%%%%%%%%%%%%%%%%%%%%%%%%%%%%%%%%%%%%%%%%%%%%%%%%%%%%%%%%%%%%%%%%%%%%%%%%%%%%%%%%%%%%%%%%%%%%%%%%%%%%%%%%%%%%%%%%%%%%%%%%%%%%%%%%%%%%%%%%%%%%%%%%%%%%%%%%%%%%%%%%%%%%%%%%%%%%%%%%%%%%%%%%%%%%%%%%%%%%%%%%%%%%%%%%%%%%%%%%%%%%%%%%%%%%%%%%%%%%%%%%%%%%%%%%%%%%%%%%%%%%%%%%%%%%%

\section{Introduction}
\label{sec:bbgw-introduction}

The observation of gravitational waves from compact binaries and the association of GW170817 with electromagnetic radiation have allowed Lorentz symmetry to be tested through the propagation of gravity~\cite{Abbott2016Observation,Abbott2017Speed,KosteleckyMewes2016,Mewes2019}. Bumblebee gravity takes into account spontaneous symmetry breaking through a vector field with a nonvanishing vacuum expectation value~\cite{KosteleckySamuel1989,Kostelecky2004,BluhmKostelecky2005,AraujoFilho:2025zaj,AraujoFilho:2025hkm}. Its nonminimal coupling to curvature relates the preferred spacetime direction to the metric dynamics, while the symmetry breaking potential controls fluctuations of the vector norm~\cite{Bluhm2008}.

The polarization content of bumblebee waves depends on the vacuum orientation and the parameters of the theory~\cite{Liang2022}. Studies of radiation in the weak field regime have derived modified wave equations, retarded Green functions, and the emission produced by localized sources~\cite{Amarilo2024}. Related calculations in metric affine bumblebee gravity and the gravitational Standard Model Extension have examined the relation between source generation, propagation, and the observed metric perturbation~\cite{AraujoFilho2026MetricAffine,AraujoFilho2026SME,AraujoFilho2026Propagation}.

The radial vacuum of Casana et al.~\cite{Casana2018} admitted an exact spherical metric with a Schwarzschild temporal component and a deformation of the radial geometry. Subsequent studies included a cosmological constant, rotation, gradient vector configurations, and more general spherical vacua~\cite{MalufNeves2021,DingChen2021,PoulisSoares2022,XuLiangShao2023} (and its corresponding non--commutative version \cite{AraujoFilho:2025rvn}). Charged solutions and recent classifications further exhibited the dependence on the vector configuration~\cite{Liu2025Charged,ZhuLi2026,Liu2026Vacua}. Likewise, the metric affine formulation admitted static and stationary axisymmetric vacuum geometries~\cite{AraujoFilho2023Vacuum,AraujoFilho2024Axisymmetric,AraujoFilho2026GeneralStatic}. Constructions with additional supporting matter also examined regular vector continuation and the characteristics of the coupled fields~\cite{AraujoFilho2026SimpsonVisser}.

Quasinormal calculations examined scalar and tensorial modes on static bumblebee geometries~\cite{Oliveira2021QNMs}, while scalar scattering and time evolution were studied in the metric affine formulation~\cite{AraujoFilho2024Traces,Heidari2024Scattering}. Analyses of slowly rotating backgrounds also included scalar, vector, and axial gravitational equations~\cite{Liu2023SlowQNMs}. On the other hand, the coupled gravitational and bumblebee equations were examined for dynamical stability~\cite{Mai2024}, and charged backgrounds were studied through coupled gravitoelectromagnetic perturbations~\cite{Li2026ChargedQNMs}.

For the static radial vacuum, Ref.~\cite{Liu2026Decoupling} recently obtained an exact decoupling of the metric and bumblebee perturbations. Under the vacuum conditions adopted in that analysis, the gravitational master equations retained the forms of Regge and Wheeler and of Zerilli, together with their parity isospectrality~\cite{ReggeWheeler1957,Zerilli1970}. In addition, the vector equations instead exhibited a coupling dependent characteristic speed, and the reconstruction of the original metric retained a dependence on the vector variables. Quasinormal frequencies locate poles of a homogeneous operator, whereas excitation amplitudes also depend on the source and boundary conditions~\cite{SasakiTagoshi2003,BertiCardosoStarinets2009,Berti2026Spectroscopy}.

Orbital calculations provide a complementary description of waveforms in bumblebee gravity. The effect of vector hair on inspiral signals was investigated in Ref.~\cite{Liang2023VectorHair}, while periodic geodesics and approximate inspiral waveforms were studied in Refs.~\cite{Shi2026PeriodicOrbits,Long2026EMRI}. The latter descriptions take into account the modified orbital motion and evaluate an adopted tensor emission prescription along the trajectories. A calculation from the field equations additionally determines which orbital stresses excite the vector sector, how that excitation propagates, and how it enters the tidal response. This distinction is particularly relevant for a neutral secondary, whose minimal worldline action supplies no direct bumblebee charge.

In this work, we derive this sourced response for an ordinary neutral point particle in the radial bumblebee exterior. We work to first order in the matter mass and expand in the nonminimal coupling with the background parameters $b$ and $\kappa$ held fixed. The particle's curvature produces an effective vector current, whose conservation also determines a response of the symmetry breaking potential. The dynamics of this potential and configurations away from its minimum have been examined in Refs.~\cite{Bluhm2008,Bailey2025Potential}. Here, the localized circular response gives $\delta V'=\xi\rho/4$, whereas radial motion introduces a conservation equation with boundary data. The axial angular current follows independently from the particle's radial angular stress.

For a slightly eccentric orbit, we obtain an oscillating $(L,m)=(1,0)$ dipole at $\Omega$ and a calculated $(2,1)$ quadrupole contribution at $2\Omega$. Their leading outgoing amplitudes scale as $ev^3$ and $ev^4$. We construct the retarded radial response, determine the leading coefficients by weak field matching, and reconstruct the dipole metric directly from its invariant equations~\cite{MartelPoisson2005,NagarRezzolla2005}. The observer's tidal tensor then gives a radial transverse strain proportional to $\xi^2b^2\mu ev^3/R$. Moreover, the Maxwell stress of the vector perturbation yields a partial dipole luminosity proportional to $\xi^2b^2\mu^2e^2v^{12}/M^2$.

The calculation refers to the specified radial exterior and its ingoing master boundary prescription. A complete binary description also involves the sourced polar equations, the effective action of a compact secondary, and the full orbital energy balance~\cite{Eardley1975,Foster2007,BarackPound2019}.

%%%%%%%%%%%%%%%%%%%%%%%%%%%%%%%%%%%%%%%%%%%%%%%%%%%%%%%%%%%%%%%%%%%%%%%%%%%%%%%%
\section{Bumblebee gravity in the presence of an orbiting source}
\label{sec:bbgw-theory}
%%%%%%%%%%%%%%%%%%%%%%%%%%%%%%%%%%%%%%%%%%%%%%%%%%%%%%%%%%%%%%%%%%%%%%%%%%%%%%%%

%%%%%%%%%%%%%%%%%%%%%%%%%%%%%%%%%%%%%%%%%%%%%%%%%%%%%%%%%%%%%%%%%%%%%%%%%%%%%%%%
\subsection{Action and radial vacuum configuration}
\label{subsec:bbgw-action}
%%%%%%%%%%%%%%%%%%%%%%%%%%%%%%%%%%%%%%%%%%%%%%%%%%%%%%%%%%%%%%%%%%%%%%%%%%%%%%%%

We use metric signature $(-,+,+,+)$, set $c=1$, and retain $\kappa=8\pi G$. The metric and bumblebee vector vary independently, while the connection is fixed to that of Levi Civita. This metric formulation differs from formulations with an independent connection~\cite{AraujoFilho2023Vacuum,AraujoFilho2024Axisymmetric}. For a neutral point particle, the action is
\begin{equation}
\begin{aligned}
S={}&\int \mathrm{d}^{4}x\sqrt{-g}\Biggl[\frac{R+\xi B^{\mu}B^{\nu}R_{\mu\nu}}{2\kappa} -\frac14F_{\mu\nu}F^{\mu\nu}-V(X)\Biggr] \\
& -\mu\int \mathrm{d}\tau,
\end{aligned}
\label{eq:bbgw-action}
\end{equation}
where $X=B_{\mu}B^{\mu}-b^{2}$, $F_{\mu\nu}=2\nabla_{[\mu}B_{\nu]}$, $b^{2}>0$, and $\mu$ is the particle mass. The worldline term couples minimally to the metric and supplies no direct bumblebee current. For a secondary black hole, internal gravitational structure can introduce a dependence of its effective action on the surrounding fields. Such effects have been studied in other compact binary theories~\cite{Eardley1975,Foster2007}.

In addition, variation with respect to $B_{\mu}$ gives
\begin{equation}
\nabla_{\mu}F^{\mu\nu}=2V'(X)B^{\nu}-\frac{\xi}{\kappa}B_{\mu}R^{\mu\nu}.
\label{eq:bbgw-vector-eom}
\end{equation}
The metric equation can be written as
\begin{equation}
\begin{aligned}
G_{\mu\nu}={}&\kappa\Bigl(T_{\mu\nu}^{\mathrm{p}}+T_{\mu\nu}^{F}+2V'B_{\mu}B_{\nu} -Vg_{\mu\nu}\Bigr)+\xi C_{\mu\nu},\\ T_{\mu\nu}^{F}={}&F_{\mu\alpha}F_{\nu}{}^{\alpha} -\frac14g_{\mu\nu}F_{\alpha\beta}F^{\alpha\beta},
\end{aligned}
\label{eq:bbgw-metric-eom}
\end{equation}
where the curvature interaction contributes
\begin{equation}
\begin{aligned}
C_{\mu\nu}={}&\frac12g_{\mu\nu}B^{\alpha}B^{\beta}R_{\alpha\beta} -B_{\mu}B^{\alpha}R_{\alpha\nu}-B_{\nu}B^{\alpha}R_{\alpha\mu}\\ &+\frac12\nabla_{\alpha}\nabla_{\mu}(B^{\alpha}B_{\nu})+\frac12\nabla_{\alpha}\nabla_{\nu}(B^{\alpha}B_{\mu})\\ &-\frac12\nabla_{\alpha}\nabla^{\alpha}(B_{\mu}B_{\nu}) -\frac12g_{\mu\nu}\nabla_{\alpha}\nabla_{\beta}(B^{\alpha}B^{\beta}).
\end{aligned}
\label{eq:bbgw-curvature-tensor}
\end{equation}
We vary the fields before imposing the vacuum norm so that it keeps the potential response to the matter stress~\cite{Bluhm2008}.

We consider the exterior vacuum metric~\cite{Casana2018,Liu2026Decoupling}
\begin{equation}
\mathrm{d}\bar s^{2} = -f(r)\mathrm{d}t^{2}+\frac{q}{f(r)}\mathrm{d}r^{2} +r^{2}\mathrm{d}\Omega^{2},
\label{eq:bbgw-background}
\end{equation}
where $f(r)=1-2M/r$, $q=1+\ell$, and $\ell=\xi b^{2}$. The radial vacuum vector is $\bar B_{\mu}\mathrm{d}x^{\mu}=b_{r}(r)\mathrm{d}r$, with $b_{r}(r)=b\sqrt{q/f(r)}$, and satisfies $\bar B_{\mu}\bar B^{\mu}=b^{2}$ and $\bar F_{\mu\nu}=0$. The mass parameter $M$ has dimensions of length and is fixed by the asymptotic circular orbital frequency. Here, $r$ is an areal radius, and we assume $q>0$ in the exterior region $r>2M$. Appendix~\ref{app:bbgw-exterior} discusses the vector case and the inner boundary prescription. The distant observer radius $R$ likewise denotes an areal radius.

%%%%%%%%%%%%%%%%%%%%%%%%%%%%%%%%%%%%%%%%%%%%%%%%%%%%%%%%%%%%%%%%%%%%%%%%%%%%%%%%%%%%%%%%%%%%%%%%%%%%%%%%%%%%%%%%%%%%%%%%%%%%%%%%%%%%%%%%%%%%%%%%%%%%%%%%%%%%%%%%%%%%%%%%%%%%%%%%%%%%%%%%%%%%%%%%%%%%%%%%%%%%%%%%%%%%%%%%%%%%%%%%%%%%%%%%%%%%%%%%%%%%%%%%%%%%%%%%%%%%%%%%%%%%%%%%%%%%%%%%%%%%%%%%%%%%%%%%%%%%%%%%%%%%%%%%%%%%%%%%%%%%%%%%%%%%%%%%%%%%%%%%%%%%%%%%%%%%%%%%%%%%%%%%%%%%%%%%%%%%%%%%%%%%%%%%%%%%%%%%%%%%%%%%%%%%%%%%%%%%%%%%%%%%%%%%%%%%%%%%%%%%%%%%%%%%%%%%%%%%%%%%%%%%%%%%%%%%%%%%%%%%%%%%%%%%%%%%%%%%%%%%%%%%%%%%%%%%%%%%%%%%%%%%%%%%%%%%%%%%%%%%%%%%%%%%%%%%%%%
\subsection{Perturbation orders and the matter stress tensor}
\label{subsec:bbgw-orders}

Now, let us write
\begin{equation}
\begin{gathered}
g_{\mu\nu}=\bar g_{\mu\nu}+h_{\mu\nu}, \,\quad B_{\mu}=\bar B_{\mu}+\beta_{\mu},\, \quad \eta=\frac{G\mu}{M}\ll1,
\end{gathered}
\label{eq:bbgw-perturbations}
\end{equation}
and retain terms linear in $\eta$ in the field equations and of order $\eta^{2}$ in the radiative energy. For the coupling expansion, we hold $b$ and $\kappa$ fixed, replace $\xi\rightarrow\epsilon\xi$, and expand in the dimensionless parameter $\epsilon$. The vector source and its explicit orbital amplitudes are calculated at first order in $\epsilon$, while the propagating axial vector contribution to the metric begins at second order in $\epsilon$. Background metric corrections and the polar potential response can appear at first order. We set $\epsilon$ to unity in the displayed physical coefficients.

The matter stress tensor is
\begin{equation}
\begin{aligned}
T_{\mathrm{p}}^{\mu\nu}(x)={}&\mu\int \frac{U^{\mu}U^{\nu}}{\sqrt{-\bar g}} \delta^{(4)}\!\left[x-z(\tau)\right]\mathrm{d}\tau,
\end{aligned}
\label{eq:bbgw-point-source}
\end{equation}
where $U^{\mu}=\mathrm{d}z^{\mu}/\mathrm{d}\tau$. For an equatorial orbit $r=r_{\mathrm{p}}(t)$, $\theta=\pi/2$, $\phi=\phi_{\mathrm{p}}(t)$, this becomes
\begin{equation}
\begin{aligned}
T_{\mathrm{p}}^{\mu\nu}= \rho_{q}U^{\mu}U^{\nu}, \quad \rho_{q}= \frac{\mu}{U^{t}\sqrt{q}\,r^{2}}\delta(r-r_{\mathrm{p}}) \times\delta_{\Omega}(\theta,\phi-\phi_{\mathrm{p}}),
\end{aligned}
\label{eq:bbgw-density-q}
\end{equation}
with $\int\delta_{\Omega}\,\mathrm{d}\Omega={}1$ and trace $T_{\mathrm{p}}=-\rho_{q}$. At first coupling order, background factors multiplying an explicit $\xi$ can be evaluated at $q=1$. We denote the resulting density by $\rho$. In other words, the matter generates the leading vector source through curvature, despite carrying no vector charge in its worldline action, which turns out to be a remarkable feature.

%%%%%%%%%%%%%%%%%%%%%%%%%%%%%%%%%%%%%%%%%%%%%%%%%%%%%%%%%%%%%%%%%%%%%%%%%%%%%%%%
\section{Conservation of the induced current and response of the potential}
\label{sec:bbgw-current}
%%%%%%%%%%%%%%%%%%%%%%%%%%%%%%%%%%%%%%%%%%%%%%%%%%%%%%%%%%%%%%%%%%%%%%%%%%%%%%%%

%%%%%%%%%%%%%%%%%%%%%%%%%%%%%%%%%%%%%%%%%%%%%%%%%%%%%%%%%%%%%%%%%%%%%%%%%%%%%%%%
\subsection{The covariant source equation}
\label{subsec:bbgw-induced-current}
%%%%%%%%%%%%%%%%%%%%%%%%%%%%%%%%%%%%%%%%%%%%%%%%%%%%%%%%%%%%%%%%%%%%%%%%%%%%%%%%

The antisymmetry of $F^{\mu\nu}$ gives
\begin{equation}
\nabla_{\nu}\nabla_{\mu}F^{\mu\nu}=\frac12[\nabla_{\nu},\nabla_{\mu}]F^{\mu\nu}=0.
\label{eq:bbgw-double-divergence}
\end{equation}
The curvature terms reduce to contractions of the symmetric Ricci tensor with $F^{\mu\nu}$. Equation~\eqref{eq:bbgw-double-divergence} requires conservation of the source in Eq.~\eqref{eq:bbgw-vector-eom}.

Let $\lambda=\delta V'$. At zero coupling, we take the field strength to vanish and choose the norm adjustment of $\beta_{\mu}$ locally as a gradient. The Maxwell stress has no term linear in the particle mass, and the curvature entering the first coupling correction is
\begin{equation}
\delta R^{\mu\nu}_{(0)}=\kappa\left(T_{\mathrm{p}}^{\mu\nu}-\frac12\bar g_{(0)}^{\mu\nu}T_{\mathrm{p}}\right).
\label{eq:bbgw-gr-curvature}
\end{equation}
The subscript $(0)$ denotes the Schwarzschild limit. The explicit factor $\xi$ in Eq.~\eqref{eq:bbgw-vector-eom} places nonminimal curvature corrections at the next coupling order, as it does a vector fluctuation multiplying the background Ricci tensor. We obtain
\begin{equation}
\begin{aligned}
\bar\nabla_{\mu}\delta F^{\mu\nu}={}&J_{\mathrm{eff}}^{\nu}, \, \, \, J_{\mathrm{eff}}^{\nu}={}&2\lambda b^{\nu}-\xi b_{\mu}T_{\mathrm{p}}^{\mu\nu} +\frac{\xi}{2}b^{\nu}T_{\mathrm{p}},
\end{aligned}
\label{eq:bbgw-induced-current}
\end{equation}
where $b^{\nu}=b\sqrt f\,\delta^{\nu}_{r}$ at the retained order. Introducing $\sigma=b_{\mu}U^{\mu}$ for the particle source gives
\begin{equation}
\begin{gathered}
J_{\mathrm{eff}}^{\nu}=b^{\nu}\left(2\lambda-\frac{\xi\rho}{2}\right)-\xi\rho\sigma U^{\nu}, \quad \bar\nabla_{\nu}J_{\mathrm{eff}}^{\nu}=0.
\end{gathered}
\label{eq:bbgw-current-dust}
\end{equation}
The first term describes the potential response along the background vector; the second depends on the projection of the orbital velocity onto that vector.

%%%%%%%%%%%%%%%%%%%%%%%%%%%%%%%%%%%%%%%%%%%%%%%%%%%%%%%%%%%%%%%%%%%%%%%%%%%%%%%%
\subsection{Circular motion and the allowed potential response}
\label{subsec:bbgw-circular-current}
%%%%%%%%%%%%%%%%%%%%%%%%%%%%%%%%%%%%%%%%%%%%%%%%%%%%%%%%%%%%%%%%%%%%%%%%%%%%%%%%

A circular geodesic at $r=r_{0}$ has
\begin{equation}
\begin{gathered}
U^{\mu}=U^{t}(1,0,0,\Omega),\quad U^{t}=\left(1-\frac{3M}{r_{0}}\right)^{-1/2},
\end{gathered}
\label{eq:bbgw-circular-geodesic}
\end{equation}
with $\Omega^{2} = M/r_{0}^{3}$ and $ \sigma=0$. Stable circular geodesics satisfy $r_{0}>6M$. The azimuthal frequency follows from $f'(r_{0})/(2r_{0})$ and is independent of $q$ at fixed $M$ and $r_{0}$. The current reduces to $J_{\mathrm{eff}}^{\nu}=b^{\nu}(2\lambda-\xi\rho/2)$.

Imposing the vacuum prescription $\lambda=0$ in the presence of the particle gives
\begin{equation}
\begin{aligned}
J_{\mathrm{eff}}^{r}={}&-\frac{\xi\mu b\sqrt f}{2U^{t}r^{2}}\delta(r-r_{0})\delta_{\Omega},\\ \bar\nabla_{\nu}J_{\mathrm{eff}}^{\nu}={}&-\frac{\xi\mu b}{2U^{t}r^{2}} \times\partial_{r}\!\left[\sqrt f\,\delta(r-r_{0})\right]\delta_{\Omega}.
\end{aligned}
\label{eq:bbgw-circular-divergence}
\end{equation}
The derivative of the radial delta distribution avoids current conservation, and notice that the obstruction persists for a smooth localized density. The regular coupling expansion needs that a potential response for this source there exists. Other vacuum configurations and matter couplings obey their own source equations. In addition, it is worth pointing out that the role of the matter interaction also appears in bumblebee solutions with nonlinear electrodynamics~\cite{LiLu2026}.

Taking into account $\lambda$ gives
\begin{equation}
\begin{gathered}
\partial_{r}\!\left[r^{2}b^{r}\left(2\lambda-\frac{\xi\rho}{2}\right)\right]=0,\qquad 2\lambda-\frac{\xi\rho}{2}=\frac{C(t,\theta,\phi)}{r^{2}b^{r}}.
\end{gathered}
\label{eq:bbgw-circular-conservation}
\end{equation}
We select a potential response localized with the matter and exclude an independent radial current through the boundaries. These conditions set $C=0$, yielding $\lambda=\delta V' = \xi\rho/4 =-\xi T_{\mathrm{p}}/4$. At first coupling order, the vector equation then admits $\delta F_{\mu\nu}=0$ with zero independent vector data. The potential stress and nonminimal metric terms can still modify the polar metric response.

For a smooth potential $V(X)=KX^{2}/2$, the same relation gives
\begin{equation}
\begin{gathered}
\delta X=\frac{\xi\rho}{4K},\qquad \delta V'=K\delta X=\frac{\xi\rho}{4}.
\end{gathered}
\label{eq:bbgw-smooth-potential}
\end{equation}
The limit of large $K$, taken after determining the response, sends $\delta X$ to zero while leaving $\delta V'$ finite. An independently varied multiplier instead imposes the norm exactly while keeping a potential response. These choices obey different field equations~\cite{Bluhm2008}; configurations away from the potential minimum have also been examined in Ref.~\cite{Bailey2025Potential}. At linear order, the particle relations are distributions. A smooth compact density gives rise to a local description before forming products of matter induced potential fluctuations. Locally, $\beta_{\mu}=\partial_{\mu}\chi$ implements the norm variation through
\begin{equation}
\delta X=2b^{r}\partial_{r}\chi-h_{\mu\nu}b^{\mu}b^{\nu}.
\label{eq:bbgw-gradient-response}
\end{equation}
Equations~\eqref{eq:bbgw-smooth-potential} and \eqref{eq:bbgw-gradient-response} determine the local vector response. Also, the global polar solution also needs that the metric equations and their boundary data be taken into account.

%%%%%%%%%%%%%%%%%%%%%%%%%%%%%%%%%%%%%%%%%%%%%%%%%%%%%%%%%%%%%%%%%%%%%%%%%%%%%%%%%%%%%%%%%%%%%%%%%%%%%%%%%%%%%%%%%%%%%%%%%%%%%%%%%%%%%%%%%%%%%%%%%%%%%%%%%%%%%%%%%%%%%%%%%%%%%%%%%%%%%%%%%%%%%%%%%%%%%%%%%%%%%%%%%%%%%%%%%%%%%%%%%%%%%%%%%%%%%%%%%%%%%%%%%%%%%%%%%%%%%%%%%%%%%%%%%%%%%%%%%%%%%%%%%%%%%%%%%%%%%%%%%%%%%%%%%%%%%%%%%%%%%%%%%%%%%%%%%%%%%%%%%%%%%%%%%%%%%%%%%%%%%%%%%%%%%%%%%%%%%%%%%%%%%%%%%%%%%%%%%%%%%%%%%%%%%%%%%%%%%%%%%%%%%%%%%%%%%%%%%%%%%%%%%%%%%%%%%%%%%%%%%%%%%%%%%%%%%%%%%%%%%%%%%%%%%%%%%%%%%%%%%%%%%%%%%%%%

\subsection{Radial motion and the radial conservation equation}
\label{subsec:bbgw-radial-current}

For a geodesic source, $\bar\nabla_{\nu}(\rho U^{\nu})=0$. Taking the divergence of Eq.~\eqref{eq:bbgw-current-dust} gives
\begin{equation}
\partial_{r}\!\left[r^{2}b^{r}\left(2\lambda-\frac{\xi\rho}{2}\right)\right]=\xi r^{2}\rho\frac{\mathrm{d}\sigma}{\mathrm{d}\tau}.
\label{eq:bbgw-radial-conservation}
\end{equation}
The general radial solution is
\begin{equation}
\begin{aligned}
2\lambda-\frac{\xi\rho}{2}={}&\frac{1}{r^{2}b^{r}}\Biggl[C_{\mathrm{in}}(t,\theta,\phi) +\xi\int_{r_{\mathrm{in}}}^{r}\mathrm{d}\tilde r\,\tilde r^{2}\rho(t,\tilde r,\theta,\phi) \\
&\times\frac{\mathrm{d}\sigma}{\mathrm{d}\tau}(t,\tilde r,\theta,\phi)\Biggr].
\end{aligned}
\label{eq:bbgw-radial-potential}
\end{equation}
Radial motion adds boundary data to the circular algebraic response and can produce a contribution extending beyond the particle's instantaneous position. This response belongs to the polar equations because $\lambda$ is a scalar perturbation. The axial angular source instead follows independently from the angular projection of the second term in Eq.~\eqref{eq:bbgw-current-dust}. Indded, a complete orbital waveform combines both contributions.

%%%%%%%%%%%%%%%%%%%%%%%%%%%%%%%%%%%%%%%%%%%%%%%%%%%%%%%%%%%%%%%%%%%%%%%%%%%%%%%%%%%%%%%%%%%%%%%%%%%%%%%%%%%%%%%%%%%%%%%%%%%%%%%%%%%%%%%%%%%%%%%%%%%%%%%%%%%%%%%%%%%%%%%%%%%%%%%%%%%%%%%%%%%%%%%%%%%%%%%%%%%%%%%%%%%%%%%%%%%%%%%%%%%%%%%%%%%%%%%%%%%%%%%%%%%%%%%%%%%%%%%%%%%%%%%%%%%%%%%%%%%%%%%%%%%%%%%%%%%%%%%%%%%%%%%%%%%%%%%%%%%%%%%%%%%%%%%%%%%%%%%%%%%%%%%%%%%%%%%%%%%%%%%%%%%%%%%%%%%%%%%%%%%%%%%%%%%%%%%%%%%%%%%%%%%%%%%%%%%%%%%%%%%%%%%%%%%%%%%%%%%%%%%%%%%%%%%%%%%%%%%%%%%%%%%%%%%%%%%%%%%%%%%%%%%%%%%%%%%%%%%%%%%%%%%%%%%%%%%%%%%%%%%%%%%%%%%%%%%%%%%%%%%%%%%%%%%%%%%%%%%%%%%%%%%%%%%%%%%%%%%%%%%%%%%%%%%%%%%%%%%%%%%%%%%%%%%%%%%%%%%%%%%%%%%%%%%%%%%%%%%%%%%%%%%%%%%%%%%%%%%%%%%%%%%%%%%%%%%%%%%%%%%%

\section{Axial perturbations and the excitation by orbital motion}
\label{sec:bbgw-axial}

\subsection{Angular harmonics and master operators}
\label{subsec:bbgw-harmonics}

Now, let us consider that $\Omega_{AB}$ and $\mathcal D_{A}$ be the metric and derivative on the unit sphere. We use normalized scalar harmonics and define the axial vector harmonics through
\begin{equation}
\begin{gathered}
\int Y_{Lm}^{*}Y_{L'm'}\,\mathrm{d}\Omega=\delta_{LL'}\delta_{mm'},\quad S_{\theta}^{Lm}=-\frac{1}{\sin\theta}\partial_{\phi}Y_{Lm},\\ S_{\phi}^{Lm}=\sin\theta\,\partial_{\theta}Y_{Lm}.
\end{gathered}
\label{eq:bbgw-harmonic-definitions}
\end{equation}
The harmonics satisfy
\begin{equation}
\begin{aligned}
{\cal D}_{A}S^{A}_{Lm}={}&0,\quad \int \Omega^{AB}S_{A}^{Lm*}S_{B}^{L'm'}\,\mathrm{d}\Omega=n_{L}\delta_{LL'}\delta_{mm'},
\end{aligned}
\label{eq:bbgw-harmonic-normalization}
\end{equation}
where $n_{L} = L(L+1)$. We expand the axial perturbations with $L\geq1$ as
\begin{equation}
\begin{gathered}
h_{tA}=\sum_{Lm}h_{0}^{Lm}S_{A}^{Lm},\qquad h_{rA}=\sum_{Lm}h_{1}^{Lm}S_{A}^{Lm},\\ \beta_{A}=\sum_{Lm}u_{Lm}S_{A}^{Lm}.
\end{gathered}
\label{eq:bbgw-axial-decomposition}
\end{equation}
For $L\geq2$, we use the Regge Wheeler gauge for the axial metric and treat the angular dipole directly in Sec.~\ref{subsec:bbgw-dipole-reconstruction}. The vacuum vector equation of Ref.~\cite{Liu2026Decoupling} is ${\cal L}_{B,L}u_{Lm}=0$, where ${\cal L}_{B,L}=\partial_{r_*}^{2}-\gamma\partial_{t}^{2}-\gamma f n_{L}/r^{2}$ and $\gamma=q-\ell^{2}/(2\kappa b^{2})$. Here, $r_*=r+2M\ln[r/(2M)-1]$ is the tortoise coordinate, and $c_{\mathrm{eff}}=\gamma^{-1/2}$ is the characteristic speed in the $(t,r_*)$ coordinates. We assume $\gamma>0$. The axial gravitational variable obeys the Schwarzschild operator
\begin{equation}
\begin{gathered}
{\cal L}_{g,L}=\partial_{r_*}^{2}-\partial_{t}^{2}-V_{\mathrm{RW},L},\qquad V_{\mathrm{RW},L}=f\left[\frac{n_{L}}{r^{2}}-\frac{6M}{r^{3}}\right].
\end{gathered}
\label{eq:bbgw-gravitational-master}
\end{equation}
The even vacuum operator has the Zerilli form~\cite{Liu2026Decoupling}. The axial equations also show the angular vector dipole, while the scalar potential response of Sec.~\ref{sec:bbgw-current} supplies no direct axial angular source at linear order.

%%%%%%%%%%%%%%%%%%%%%%%%%%%%%%%%%%%%%%%%%%%%%%%%%%%%%%%%%%%%%%%%%%%%%%%%%%%%%%%%
\subsection{Projection of the matter source}
\label{subsec:bbgw-source-projection}
%%%%%%%%%%%%%%%%%%%%%%%%%%%%%%%%%%%%%%%%%%%%%%%%%%%%%%%%%%%%%%%%%%%%%%%%%%%%%%%%

At first order in the coupling, Eq.~\eqref{eq:bbgw-induced-current} gives $J_{\mathrm{eff}}^{A}=-\xi b_{r}T_{\mathrm{p}}^{rA}$ in the axial sector. We define
\begin{equation}
\begin{gathered}
\left(T_{rA}^{\mathrm{p}}\right)_{\mathrm{axial}}=\sum_{Lm}\tau_{r}^{Lm}S_{A}^{Lm},\quad \tau_{r}^{Lm}=\frac{1}{n_{L}}\int S^{A,Lm*}T_{rA}^{\mathrm{p}}\,\mathrm{d}\Omega.
\end{gathered}
\label{eq:bbgw-tau-definition}
\end{equation}
For a single vector harmonic, the angular part of the Maxwell operator on Schwarzschild is
\begin{equation}
\bar\nabla_{\mu}\delta F^{\mu A}=\frac{S^{A,Lm}}{r^{2}f}\left[\partial_{r_*}^{2}-\partial_{t}^{2}-\frac{fn_{L}}{r^{2}}\right]u_{Lm}.
\label{eq:bbgw-maxwell-projection}
\end{equation}
Since $T^{rA}=f\tau_{r}^{Lm}S^{A,Lm}/r^{2}$ at this order, the sourced equation is
\begin{equation}
\left[\partial_{r_*}^{2}-\partial_{t}^{2}-\frac{fn_{L}}{r^{2}}\right]u_{Lm}=-\xi b_{r}f^{2}\tau_{r}^{Lm}.
\label{eq:bbgw-sourced-axial}
\end{equation}
We evaluate the background factors on the right hand side at zero coupling. The covariant vector equation and the angular norm in Eq.~\eqref{eq:bbgw-harmonic-normalization} fix the normalization of Eq.~\eqref{eq:bbgw-sourced-axial}.

For the equatorial orbit, let ${\cal L}_{\mathrm{p}}=U_{\phi}=r_{\mathrm{p}}^{2}U^{\phi}$ denote the specific angular momentum. The source projection becomes
\begin{equation}
\begin{aligned}
\tau_{r}^{Lm}={}&\frac{\mu U_{r}{\cal L}_{\mathrm{p}}}{n_{L}U^{t}r_{\mathrm{p}}^{2}}\delta(r-r_{\mathrm{p}}) S^{\phi,Lm*}(\pi/2,\phi_{\mathrm{p}}).
\end{aligned}
\label{eq:bbgw-tau-particle}
\end{equation}
Radial motion transports angular momentum across surfaces of fixed areal radius. The resulting stress couples to the radial vacuum vector and produces an angular current. The equatorial factor $\partial_\theta Y_{Lm}(\pi/2)$ selects $L+m$ odd, including $(L,m)=(1,0)$ and $(2,\pm1)$. For circular motion, $U_{r}=0$, so the axial vector source in Eq.~\eqref{eq:bbgw-sourced-axial} vanishes. Other stress tensor projections still excite gravitational multipoles.

%%%%%%%%%%%%%%%%%%%%%%%%%%%%%%%%%%%%%%%%%%%%%%%%%%%%%%%%%%%%%%%%%%%%%%%%%%%%%%%%
\subsection{Metric reconstruction and the angular dipole}
\label{subsec:bbgw-dipole-reconstruction}
%%%%%%%%%%%%%%%%%%%%%%%%%%%%%%%%%%%%%%%%%%%%%%%%%%%%%%%%%%%%%%%%%%%%%%%%%%%%%%%%

For $L\geq2$, define $D_{L}=n_{L}-2$. The vacuum reconstruction of Ref.~\cite{Liu2026Decoupling} contains
\begin{equation}
\begin{aligned}
h_{0}^{Lm}&=-\frac{2f}{D_{L}}\left(\Psi_{g}^{Lm}+r\partial_{r}\Psi_{g}^{Lm}\right),\\ h_{1}^{Lm}&=-\frac{2r}{D_{L}f}\partial_{t}\Psi_{g}^{Lm}+\frac{\ell}{fb_{r}}u_{Lm}.
\end{aligned}
\label{eq:bbgw-metric-reconstruction}
\end{equation}
The last term gives the vector contribution to the original metric. The angular dipole has $D_{1}=0$ and needs its own equations~\cite{MartelPoisson2005,NagarRezzolla2005}. We introduce the axial invariant
\begin{equation}
{\cal F}_{1}=\partial_{r}h_{0}^{10}-\frac{2}{r}h_{0}^{10}-\partial_{t}h_{1}^{10}.
\label{eq:bbgw-dipole-invariant}
\end{equation}
Under an axial coordinate transformation with parameter $\Lambda(t,r)$, the metric functions change as
\begin{equation}
\begin{gathered}
h_{0}^{10}\longrightarrow h_{0}^{10}-\partial_{t}\Lambda,\qquad h_{1}^{10}\longrightarrow h_{1}^{10}-\partial_{r}\Lambda+\frac{2}{r}\Lambda,
\end{gathered}
\label{eq:bbgw-dipole-gauge}
\end{equation}
which leaves ${\cal F}_{1}$ invariant. In the vacuum observer region, the radial angular Einstein equation for $L=1$ gives
\begin{equation}
\partial_{t}\left[{\cal F}_{1}+\frac{\ell}{fb_{r}}\partial_{t}u_{10}\right]=0.
\label{eq:bbgw-dipole-metric-equation}
\end{equation}
The time independent integration term describes a rotational perturbation. For the oscillating vector contribution, we set $h_{0}^{10}=0$ and we obtain
\begin{equation}
\begin{gathered}
h_{1}^{10}=\frac{\ell}{fb_{r}}u_{10},\qquad {\cal F}_{1}=-\frac{\ell}{fb_{r}}\partial_{t}u_{10}.
\end{gathered}
\label{eq:bbgw-dipole-reconstruction}
\end{equation}
The dipole reconstruction avoids division by $D_{1}$ and yields a nonzero invariant ${\cal F}_{1}$ for the oscillating vector mode. A stationary angular momentum perturbation may be added independently as well. Appendix~\ref{app:bbgw-sign} specifies the relative derivative sign in the axial equations.

%%%%%%%%%%%%%%%%%%%%%%%%%%%%%%%%%%%%%%%%%%%%%%%%%%%%%%%%%%%%%%%%%%%%%%%%%%%%%%%%%%%%%%%%%%%%%%%%%%%%%%%%%%%%%%%%%%%%%%%%%%%%%%%%%%%%%%%%%%%%%%%%%%%%%%%%%%%%%%%%%%%%%%%%%%%%%%%%%%%%%%%%%%%%%%%%%%%%%%%%%%%%%%%%%%%%%%%%%%%%%%%%%%%%%%%%%%%%%%%%%%%%%%%%%%%%%%%%%%%%%%%%%%%%%%%%%%%%%%%%%%%%%%%%%%%%%%%%%%%%%%%%%%%%%%%%%%%%%%%%%%%%%%%%%%%%%%%%%%%%%%%%%%%%%%%%%%%%%%%%%%%%%%%%%%%%%%%%%%%%%%%%%%%%%%%%%%%%%%%%%%%%%%%%%%%%%%%%%%%%%%%%%%%%%%%%%%%%%%%%%%%%%%%%%%%%%%%%%%%%%%%%%%%%%%%%%%%%%%%%%%%%%%%%%%%%%%%%%%%%%%%%%%%%%%%%%%%%
\section{Retarded Green function and the radiation region}
\label{sec:bbgw-green}

\subsection{Radial response and outgoing amplitudes}
\label{subsec:bbgw-green-radial}

For a Fourier component proportional to $e^{-i\omega t}$, the exterior vacuum vector equation has radial operator
\begin{equation}
{\cal D}_{B,L}(\omega)=\frac{\mathrm{d}^{2}}{\mathrm{d}r_*^{2}}+\gamma\omega^{2}-\gamma\frac{fn_{L}}{r^{2}}.
\label{eq:bbgw-frequency-operator}
\end{equation}
We choose homogeneous solutions $X_{L\omega}^{\mathrm{in}}$ and $X_{L\omega}^{\mathrm{up}}$ with ingoing behavior at the inner master boundary and unit outgoing amplitude at large radius,
\begin{equation}
\begin{gathered}
X_{L\omega}^{\mathrm{in}}\sim e^{-ik r_*},\,\,\,\,\, X_{L\omega}^{\mathrm{up}}\sim e^{ik r_*},\,\,\,\,\, k=\sqrt{\gamma}\,\omega=\frac{\omega}{c_{\mathrm{eff}}}.
\end{gathered}
\label{eq:bbgw-green-boundaries}
\end{equation}
Their Wronskian is
\begin{equation}
W_{L}(\omega)=X_{L\omega}^{\mathrm{in}}\frac{\mathrm{d}X_{L\omega}^{\mathrm{up}}}{\mathrm{d}r_*}-X_{L\omega}^{\mathrm{up}}\frac{\mathrm{d}X_{L\omega}^{\mathrm{in}}}{\mathrm{d}r_*},
\label{eq:bbgw-wronskian}
\end{equation}
which is independent of $r_*$. The radial Green function and its source integral are
\begin{equation}
\begin{aligned}
\widetilde G_{L}(\omega;r_*,r_*')={}&\frac{X_{L\omega}^{\mathrm{in}}(r_{*<})X_{L\omega}^{\mathrm{up}}(r_{*>})}{W_{L}(\omega)},\\ X_{Lm\omega}(r)={}&\int \widetilde G_{L}(\omega;r,r')  S_{Lm\omega}(r')\,\mathrm{d}r_*'.
\end{aligned}
\label{eq:bbgw-green-function}
\end{equation}
The inverse Fourier integral uses the retarded prescription with no independent incoming vector radiation. At finite coupling, $S_{Lm\omega}$ must be obtained by reducing the full sourced system. The explicit calculations below set $\gamma=1$ and use the first order source derived in Eq.~\eqref{eq:bbgw-sourced-axial}. For a periodic source, each nonzero harmonic produces an outgoing coefficient
\begin{equation}
Z_{Lm\omega}^{\infty}=\frac{1}{W_{L}(\omega)}\int X_{L\omega}^{\mathrm{in}}(r')S_{Lm\omega}(r')\,\mathrm{d}r_*'.
\label{eq:bbgw-outgoing-coefficient}
\end{equation}
The retarded response takes into account curvature scattering and the prescribed inner absorption~\cite{SasakiTagoshi2003}. Its quasinormal poles arise from zeros of the Wronskian, while the source integral fixes their excitation. Since the angular potential in Eq.~\eqref{eq:bbgw-frequency-operator} also depends on $\gamma$, the vector spectrum involves more than a uniform frequency rescaling~\cite{Liu2026Decoupling}.

%%%%%%%%%%%%%%%%%%%%%%%%%%%%%%%%%%%%%%%%%%%%%%%%%%%%%%%%%%%%%%%%%%%%%%%%%%%%%%%%
\subsection{Leading matching for a slowly moving source}
\label{subsec:bbgw-matching}
%%%%%%%%%%%%%%%%%%%%%%%%%%%%%%%%%%%%%%%%%%%%%%%%%%%%%%%%%%%%%%%%%%%%%%%%%%%%%%%%

We calculate the explicit amplitudes for $v^{2}=M/a\ll1$ at the first coupling order of Eq.~\eqref{eq:bbgw-sourced-axial}. The leading radial equation is
\begin{equation}
\left[\frac{\mathrm{d}^{2}}{\mathrm{d}r^{2}}+\omega^{2}-\frac{n_{L}}{r^{2}}\right]X_{Lm\omega}=S_{Lm\omega}.
\label{eq:bbgw-flat-radial}
\end{equation}
The regular solution and a solution with unit outgoing amplitude are
\begin{equation}
\begin{gathered}
X_{L\omega}^{\mathrm{reg}}=zj_{L}(z),\,\,\,\, X_{L\omega}^{\mathrm{up}}=i^{L+1}zh_{L}^{(1)}(z),\\ z=\omega r,\,\,\,\, W_{L}=i^{L+2}\omega.
\end{gathered}
\label{eq:bbgw-bessel-solutions}
\end{equation}
The spherical Bessel Wronskian fixes both the normalization and phases of the dipole and quadrupole coefficients.

The static solutions provide a check of the matching to the black hole exterior. The first two axial Maxwell equations admit
\begin{equation}
\begin{aligned}
X_{1}^{\mathrm{stat}}={}r^{2},\,\,\,\,\,\, X_{2}^{\mathrm{stat}}={}r^{3}-\frac32Mr^{2},\\ \partial_{r}\!\left(f\partial_{r}X_{L}^{\mathrm{stat}}\right)-\frac{n_{L}}{r^{2}}X_{L}^{\mathrm{stat}}={}&0.
\end{aligned}
\label{eq:bbgw-static-matching}
\end{equation}
As master functions, these solutions are finite at $r=2M$. In the overlap region $2M\ll r\sim a\ll\Omega^{-1}$, they approach the regular flat behaviors $r^{L+1}$, with a relative quadrupole correction of order $M/a$. Their normalization cancels between the numerator and the Wronskian in Eq.~\eqref{eq:bbgw-outgoing-coefficient}. The resulting matching fixes the leading amplitudes.

%%%%%%%%%%%%%%%%%%%%%%%%%%%%%%%%%%%%%%%%%%%%%%%%%%%%%%%%%%%%%%%%%%%%%%%%%%%%%%%%%%%%%%%%%%%%%%%%%%%%%%%%%%%%%%%%%%%%%%%%%%%%%%%%%%%%%%%%%%%%%%%%%%%%%%%%%%%%%%%%%%%%%%%%%%%%%%%%%%%%%%%%%%%%%%%%%%%%%%%%%%%%%%%%%%%%%%%%%%%%%%%%%%%%%%%%%%%%%%%%%%%%%%%%%%%%%%%%%%%%%%%%%%%%%%%%%%%%%%%%%%%%%%%%%%%%%%%%%%%%%%%%%%%%%%%%%%%%%%%%%%%%%%%%%%%%%%%%%%%%%%%%%%%%%%%%%%%%%%%%%%%%%%%%%%%%%%%%%%%%%%%%%%%%%%%%%%%%%%%%%%%%%%%%%%%%%%%%%%%%%%%%%%%%%%%%%%%%%%%%%%%%%%%%%%%%%%%%%%%%%%%%%%%%%%%%%%%%%%%%%%%%%%%%%%%%%%%%%%%%%%%%%%%%%%%%%%%%%%%%%%%%%%%%%%%%%%%%%%%%%%%%%%%%%%%%%%%%%%%

\section{Vector radiation from a slightly eccentric orbit}
\label{sec:bbgw-eccentric}

\subsection{Orbital expansion and harmonic content}
\label{subsec:bbgw-orbit}

For a Kepler orbit with small eccentricity, we use
\begin{equation}
\begin{gathered}
r_{\mathrm{p}}=a\left[1-e\cos\Omega t+O(e^{2})\right],\\ \phi_{\mathrm{p}}=\Omega t+2e\sin\Omega t+O(e^{2}),\qquad \Omega^{2}=\frac{M}{a^{3}},
\end{gathered}
\label{eq:bbgw-kepler-orbit}
\end{equation}
\begin{equation}
\begin{aligned}
U_{r}={}&ae\Omega\sin\Omega t\left[1+O(v^{2})\right] +O(e^{2}v),\\ {\cal L}_{\mathrm{p}}={}&a^{2}\Omega\left[1+O(v^{2})+O(e^{2})\right],\qquad U^{t}={}&1+O(v^{2}).
\end{aligned}
\label{eq:bbgw-kepler-velocities}
\end{equation}
The source in Eq.~\eqref{eq:bbgw-tau-particle} is linear in $e$ through $U_{r}$, so its remaining factors may be evaluated at $r=a$ and $\phi=\Omega t$ to first order in eccentricity. We hold the orbital parameters fixed over times short compared with the radiation reaction time~\cite{PetersMathews1963,Peters1964}. Their evolution on inspiral times follows from the corresponding energy and angular momentum losses indeed.

%%%%%%%%%%%%%%%%%%%%%%%%%%%%%%%%%%%%%%%%%%%%%%%%%%%%%%%%%%%%%%%%%%%%%%%%%%%%%%%%%%%%%%%%%%%%%%%%%%%%%%%%%%%%%%%%%%%%%%%%%%%%%%%%%%%%%%%%%%%%%%%%%%%%%%%%%%%%%%%%%%%%%%%%%%%%%%%%%%%%%%%%%%%%%%%%%%%%%%%%%%%%%%%%%%%%%%%%%%%%%%%%%%%%%%%%%%%%%%%%%%%%%%%%%%%%%%%%%%%%%%%%%%%%%%%%%%%%%%%%%%%%%%%%%%%%%%%%%%%%%%%%%%%%%%%%%%%%%%%%%%%%%%%%%%%%%%%%%%%%%%%%%%%%%%%%%%%%%%%%%%%%%%%%%%%%%%%%%%%%%%%%%%%%%%%%%%%%%%%%%%%%%%%%%%%%%%%%%%%%%%%%%%%%%%%%%%%%%%%%%%%%%%%%%%%%%%%%%%%%%%%%%%%%%%%%%%%%%%%%%%%%%%%%%%%%%%%%%%%%%%%%%%%%%%%%%%%%%%%%%%%%%%%%%%%%%%%%%%%%%%%%%%%%%%%%%%%%%%%%%%%%%%%%%%%%%%%%%%%%%%%%%%%%%%%%%%%%%%%%%%

\subsection{The axial dipole}
\label{subsec:bbgw-dipole-amplitude}

For $L=1$ and $m=0$, we have 
\begin{equation}
\begin{gathered}
Y_{10}=C_{1}\cos\theta,\,\,\, C_{1}=\sqrt{\frac{3}{4\pi}},\,\,\, S_{\phi}^{10}=-C_{1}\sin^{2}\theta,\,\,\, S_{\theta}^{10}=0.
\end{gathered}
\label{eq:bbgw-dipole-harmonic}
\end{equation}
Equation~\eqref{eq:bbgw-tau-particle} gives
\begin{equation}
\tau_{r}^{10}=-\frac{\mu e a\Omega^{2}C_{1}}{2}\sin\Omega t\,\delta(r-a),
\label{eq:bbgw-dipole-stress}
\end{equation}
and the leading vector equation becomes
\begin{equation}
\begin{aligned}
\left[\partial_{r}^{2}-\partial_{t}^{2}-\frac{2}{r^{2}}\right]u_{10}={}&A_{1}\sin\Omega t\,\delta(r-a),
\end{aligned}
\label{eq:bbgw-dipole-source}
\end{equation}
where $A_{1} = \xi\, b\mu\, e \,a\,\Omega^{2}C_{1}/2$. The azimuthal number $m=0$ specifies the angular symmetry, while radial motion supplies an oscillation at frequency $\Omega$. With the convention $e^{-i\omega t}$, the positive frequency source has coefficient $s_{1,\Omega}=iA_{1}/2$. For $L=1$, Eq.~\eqref{eq:bbgw-bessel-solutions} gives $W_{1}=-i\Omega$. The outgoing amplitude reads
\begin{equation}
\begin{aligned}
Z_{10,\Omega}^{\infty}={}\frac{s_{1,\Omega}}{-i\Omega}\,\Omega a\,j_{1}(\Omega a) = -\frac{A_{1}a}{2}j_{1}(v),\quad v={}&\Omega a=\sqrt{\frac Ma}.
\end{aligned}
\label{eq:bbgw-dipole-z}
\end{equation}
Using $j_{1}(v)=v/3+O(v^{3})$, we obtain
\begin{equation}
\begin{aligned}
Z_{10,\Omega}^{\infty}={}&-\frac{\xi b\mu e C_{1}}{12}v^{3} \left[1+O(v^{2})+O(e)+O(\epsilon)\right].
\end{aligned}
\label{eq:bbgw-dipole-leading-z}
\end{equation}
The remainders are relative corrections, with $\epsilon$ denoting the higher coupling orders defined in Sec.~\ref{subsec:bbgw-orders}. The real $m=0$ field contains both temporal frequencies, giving the radiative part
\begin{equation}
\begin{aligned}
u_{10}^{\mathrm{out}}(t,R)={}&2\operatorname{Re}\!\left[Z_{10,\Omega}^{\infty}e^{-i\Omega(t-R)}\right]\\ ={}&-\frac{\xi b\mu e C_{1}}{6}v^{3} \cos[\Omega(t-R)]+\cdots.
\end{aligned}
\label{eq:bbgw-dipole-real-field}
\end{equation}
The temporal conjugate supplies the factor of two. The resulting axial dipole is linear in eccentricity and vanishes for a circular orbit. Its phase in Eq.~\eqref{eq:bbgw-dipole-real-field} is evaluated at the same coupling and weak field orders as the amplitude.

%%%%%%%%%%%%%%%%%%%%%%%%%%%%%%%%%%%%%%%%%%%%%%%%%%%%%%%%%%%%%%%%%%%%%%%%%%%%%%%%
\subsection{The axial quadrupole}
\label{subsec:bbgw-quadrupole-amplitude}
%%%%%%%%%%%%%%%%%%%%%%%%%%%%%%%%%%%%%%%%%%%%%%%%%%%%%%%%%%%%%%%%%%%%%%%%%%%%%%%%

For comparison, consider $L=2$, $m=1$, with
\begin{equation}
\begin{gathered}
Y_{21}=-C_{2}\sin\theta\cos\theta\,e^{i\phi},\quad C_{2}=\sqrt{\frac{15}{8\pi}},\\ S_{\phi}^{21}(\pi/2,\phi)=C_{2}e^{i\phi}.
\end{gathered}
\label{eq:bbgw-quadrupole-harmonic}
\end{equation}
Its source equation is
\begin{equation}
\begin{aligned}
\left[\partial_{r}^{2}-\partial_{t}^{2}-\frac{6}{r^{2}}\right]u_{21}={}&-A_{2}\sin\Omega t e^{-i\Omega t}\delta(r-a),
\end{aligned}
\label{eq:bbgw-quadrupole-source}
\end{equation}
where $ A_{2} = \xi \,b\,\mu\, e\, a\,\Omega^{2}C_{2}/6$. The time dependence separates as
\begin{equation}
-A_{2}\sin\Omega t\,e^{-i\Omega t}=\frac{iA_{2}}{2}-\frac{iA_{2}}{2}e^{-2i\Omega t}.
\label{eq:bbgw-quadrupole-frequency}
\end{equation}
The first term is stationary, whereas the second radiates at $\omega=2\Omega$ with coefficient $s_{2,2\Omega}=-iA_{2}/2$. Since $W_{2}=\omega$,
\begin{equation}
\begin{aligned}
Z_{21,2\Omega}^{\infty}={}&s_{2,2\Omega}a\,j_{2}(2v) =-\frac{i\xi b\mu e C_{2}}{45}v^{4}\\ &\times\left[1+O(v^{2})+O(e)+O(\epsilon)\right],
\end{aligned}
\label{eq:bbgw-quadrupole-z}
\end{equation}
where we used $j_{2}(z)=z^{2}/15+O(z^{4})$. The real vector field combines this coefficient with its angular and temporal conjugate. The dipole coefficient in Eq.~\eqref{eq:bbgw-dipole-leading-z} contains one fewer power of $v$ than this quadrupole. These are the two axial contributions kept here; the polar response contributes to the total radiation.

\begin{figure}[tbp]
\centering
\includegraphics[width=\columnwidth]{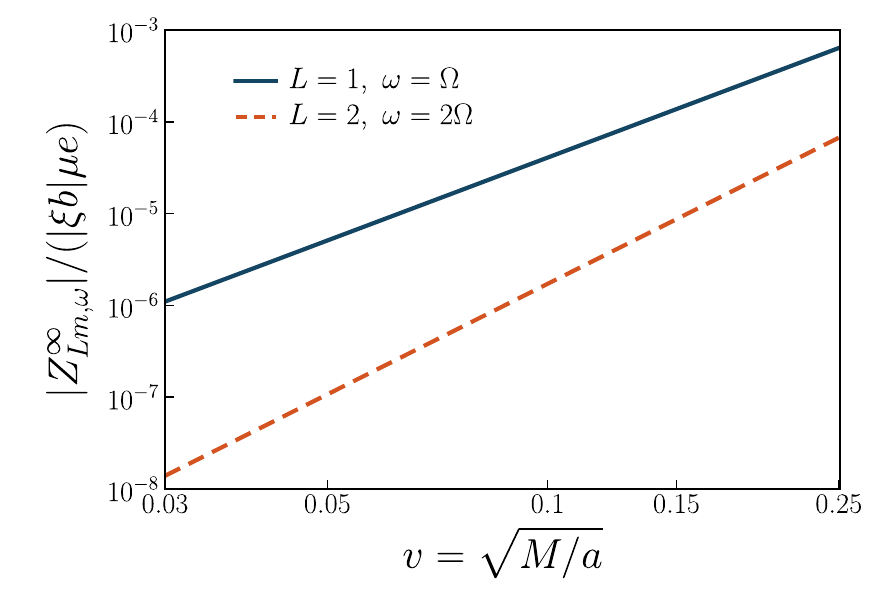}
\caption{Positive frequency outgoing coefficients for the axial dipole $(L,m,\omega)=(1,0,\Omega)$ and quadrupole $(2,1,2\Omega)$, divided by $|\xi b|\mu e$. The curves show $C_1v^3/12$ and $C_2v^4/45$, with $C_1=\sqrt{3/(4\pi)}$ and $C_2=\sqrt{15/(8\pi)}$, over $0.03\leq v\leq0.25$, where $v=\sqrt{M/a}$.}
\label{fig:bb-harmonic-amplitudes}
\end{figure}

The $v^3$ dipole amplitude exceeds the $v^4$ quadrupole amplitude throughout the interval shown in Fig.~\ref{fig:bb-harmonic-amplitudes}, with $|Z^{\infty}_{21,2\Omega}|/|Z^{\infty}_{10,\Omega}|=(4/15)\sqrt{5/2}\,v$ at the leading order. Real strain and power comparisons also take into account the angular projection and conjugate terms.

%%%%%%%%%%%%%%%%%%%%%%%%%%%%%%%%%%%%%%%%%%%%%%%%%%%%%%%%%%%%%%%%%%%%%%%%%%%%%%%%
\section{Tidal strain measured by a distant observer}
\label{sec:bbgw-strain}
%%%%%%%%%%%%%%%%%%%%%%%%%%%%%%%%%%%%%%%%%%%%%%%%%%%%%%%%%%%%%%%%%%%%%%%%%%%%%%%%

%%%%%%%%%%%%%%%%%%%%%%%%%%%%%%%%%%%%%%%%%%%%%%%%%%%%%%%%%%%%%%%%%%%%%%%%%%%%%%%%
\subsection{Curvature of the vector contribution to the metric}
\label{subsec:bbgw-tidal-curvature}
%%%%%%%%%%%%%%%%%%%%%%%%%%%%%%%%%%%%%%%%%%%%%%%%%%%%%%%%%%%%%%%%%%%%%%%%%%%%%%%%

The metric reconstruction in Eq.~\eqref{eq:bbgw-metric-reconstruction}, or Eq.~\eqref{eq:bbgw-dipole-reconstruction} for the dipole, converts the vector amplitude into a metric perturbation. To isolate its contribution in the vacuum observer region, we set the independent gravitational wave to zero and write, for a single harmonic,
\begin{equation}
\begin{gathered}
h_{tA}^{V}=0,\qquad h_{rA}^{V}=H_{Lm}S_{A}^{Lm},\qquad h_{AB}^{V}=0,\\ H_{Lm}=\frac{\ell}{fb_{r}}u_{Lm}=\frac{\ell}{b\sqrt{qf}}u_{Lm}.
\end{gathered}
\label{eq:bbgw-vector-metric}
\end{equation}
We define the strain through the relative acceleration of freely falling masses. Let $\mathcal E_{\hat i\hat j}=R_{\hat0\hat i\hat0\hat j}$ denote the electric part of the Riemann tensor in the observer's orthonormal frame. With our curvature convention, geodesic deviation obeys $\mathrm{d}^{2}\zeta^{\hat i}/\mathrm{d}\tau_{\mathrm{o}}^{2}=-\mathcal E^{\hat i}{}_{\hat j}\zeta^{\hat j}$ and a transverse metric wave gives $\delta\mathcal E_{\hat i\hat j}=-\partial_{\tau_{\mathrm{o}}}^{2}h_{\hat i\hat j}/2$. This local definition also applies to modes whose characteristic speed differs from that of light~\cite{Eardley1973,Liang2022}.

Direct evaluation of Eq.~\eqref{eq:bbgw-vector-metric} gives
\begin{equation}
\begin{aligned}
\delta R_{trtA} = {}&\left[-\frac12\partial_{t}^{2}H_{Lm}+\frac{ff'}{2qr}H_{Lm}\right]S_{A}^{Lm},\\ \delta R_{tAtB}={}&-\frac{ff'}{2q}H_{Lm}S_{AB}^{Lm},\qquad S_{AB}^{Lm}=\mathcal D_{(A}S_{B)}^{Lm}.
\end{aligned}
\label{eq:bbgw-coordinate-curvature}
\end{equation}
Orthonormality must be imposed on the perturbed spatial metric. The angular frame vectors constructed from the unit sphere dyad $\mathsf e_{\hat A}{}^{B}$ acquire a radial component proportional to $H_{Lm}$. The frame in Appendix~\ref{app:bbgw-curvature} gives
\begin{equation}
\begin{aligned}
\delta\mathcal E_{\hat r\hat A}={}&\Biggl[-\frac{\ell}{2bqfr}\partial_{t}^{2}u_{Lm} +\frac{3\ell M}{bq^{2}r^{4}}u_{Lm}\Biggr]S_{\hat A}^{Lm},\\ \delta\mathcal E_{\hat A\hat B}={}&-\frac{f'H_{Lm}}{2qr^{2}}S_{\hat A\hat B}^{Lm},
\end{aligned}
\label{eq:bbgw-orthonormal-curvature}
\end{equation}
where $S_{\hat A}=\mathsf e_{\hat A}{}^{B}S_{B}$. At large radius, the oscillating radial angular component becomes
\begin{equation}
\begin{aligned}
\delta\mathcal E_{\hat r\hat A}^{\mathrm{rad}}={}&-\frac{\ell}{2bqR}\partial_{t}^{2}u_{Lm}^{\mathrm{out}}S_{\hat A}^{Lm} +O(R^{-2}).
\end{aligned}
\label{eq:bbgw-radiative-curvature}
\end{equation}
The angular tidal term in Eq.~\eqref{eq:bbgw-orthonormal-curvature} decays as $M/R^{4}$ for fixed outgoing master amplitude, leaving a leading signal with one radial and one transverse index. After subtracting the stationary background tide, the effective strain for short detector arms at frequencies $\omega^{2}\gg M/R^{3}$ is
\begin{equation}
h_{\hat r\hat A}^{V}(t,R)=\frac{\ell}{bqR}\sum_{Lm}u_{Lm}^{\mathrm{out}}(t,R)S_{\hat A}^{Lm}.
\label{eq:bbgw-general-vector-strain}
\end{equation}
For a periodic signal, we set the integration constants describing nonoscillating displacements to zero. The radiative $1/R$ term characterizes the additional polarization. Although $S_{AB}^{10}=0$, the dipole retains $S_{\hat\phi}^{10}=-C_{1}\sin\theta$ and produces radial transverse relative motion, while general relativity has no radiative tensor dipole.

%%%%%%%%%%%%%%%%%%%%%%%%%%%%%%%%%%%%%%%%%%%%%%%%%%%%%%%%%%%%%%%%%%%%%%%%%%%%%%%%
\subsection{The explicit dipole strain and the detector projection}
\label{subsec:bbgw-dipole-strain}
%%%%%%%%%%%%%%%%%%%%%%%%%%%%%%%%%%%%%%%%%%%%%%%%%%%%%%%%%%%%%%%%%%%%%%%%%%%%%%%%

Combining Eqs.~\eqref{eq:bbgw-dipole-real-field} and \eqref{eq:bbgw-general-vector-strain} at leading coupling order gives
\begin{equation}
\begin{aligned}
h_{\hat r\hat\phi}^{V,10}(t,R)={}&\frac{\xi^{2}b^{2}\mu e}{8\pi R}v^{3}\sin\theta \,\cos[\Omega(t-R)]+\cdots,
\end{aligned}
\label{eq:bbgw-final-dipole-strain}
\end{equation}
with $h_{\hat r\hat\theta}^{V,10}=0$. The coefficient follows from $C_{1}^{2}=3/(4\pi)$ and the two temporal frequencies of the real $m=0$ field. Corrections enter at higher coupling and eccentricity orders and at relative weak field order $v^{2}$; curvature scattering also modifies the propagation phase. The strain scales linearly with the matter mass and inversely with distance, vanishes as $e\rightarrow0$, and peaks in the orbital plane with a node on the orbital axis.

The same reconstruction gives the calculated quadrupole contribution
\begin{equation}
\begin{aligned}
h_{\hat r\hat A}^{V,21}={}&-\frac{i\xi^{2}b^{2}\mu e C_{2}}{45R}v^{4} \times e^{-2i\Omega(t-R)}S_{\hat A}^{21}\\ &+\text{complex conjugate}+\cdots.
\end{aligned}
\label{eq:bbgw-final-quadrupole-strain}
\end{equation}
Adding the conjugate gives the real field. The orbital stress in Eq.~\eqref{eq:bbgw-tau-particle} fixes the amplitudes and harmonic content of the two multipoles.

\begin{figure}[tbp]
\centering
\includegraphics[width=\columnwidth]{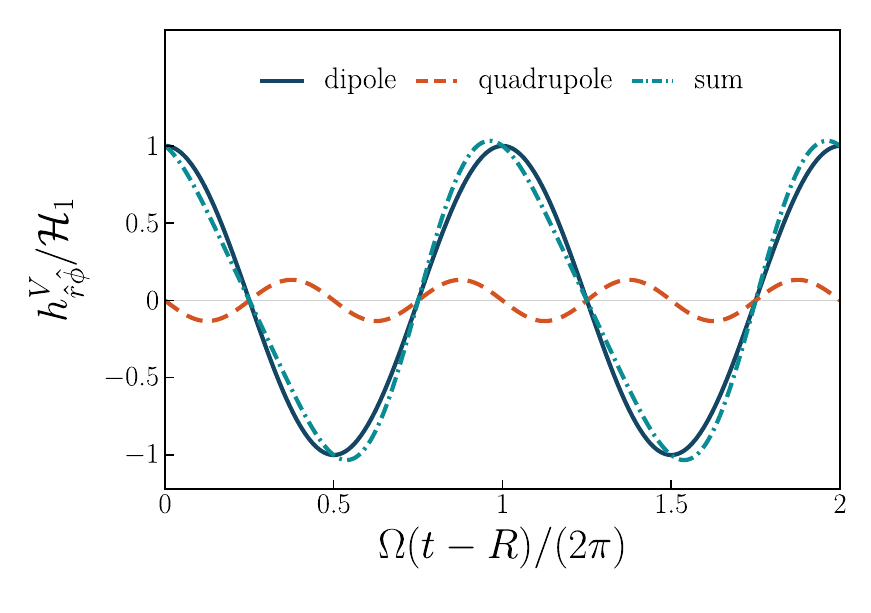}
\caption{Real radial transverse strain from the two calculated axial multipoles at $\theta=\pi/2$, $\phi=0$, and $v=0.20$. With $x=\Omega(t-R)/(2\pi)$, the curves are $h_{\hat r\hat\phi}^{V,10}/\mathcal{H}_1=\cos(2\pi x)$ and $h_{\hat r\hat\phi}^{V,21}/\mathcal{H}_1=-(2v/3)\sin(4\pi x)$, together with their sum. The real quadrupole includes its conjugate harmonic. The normalization $\mathcal{H}_1=\xi^2b^2\mu ev^3/(8\pi R)$ assumes $e>0$; both physical contributions vanish as $e\to0$. }
\label{fig:bb-axial-waveform}
\end{figure}

Figure~\ref{fig:bb-axial-waveform} compares the dipole at the orbital frequency with the quadrupole at twice that frequency. In the orbital plane at $\phi=0$, their amplitude ratio is $2v/3$ at the retained order. Their sum distorts the dipole oscillation; both physical contributions maintain linear in $e$ and vanish in the circular limit.

Let the unit vectors $\boldsymbol p$ and $\boldsymbol s$ specify the detector arm directions in a local freely falling frame, with arm lengths small compared with the wavelength. The response tensor is
\begin{equation}
\begin{gathered}
D^{\hat i\hat j}=\frac12\left(p^{\hat i}p^{\hat j}-s^{\hat i}s^{\hat j}\right),\qquad h_{\mathrm{D}}^{V}=2D^{\hat r\hat A}h_{\hat r\hat A}^{V}.
\end{gathered}
\label{eq:bbgw-detector-tensor}
\end{equation}
For the dipole, define $F_{V}=2D^{\hat r\hat\phi}$ and $\mathcal H_{1}=\xi^{2}b^{2}\mu e v^{3}/(8\pi R)$. Then, we have
\begin{equation}
h_{\mathrm{D}}^{V,10}=F_{V}\mathcal H_{1}\sin\theta\cos[\Omega(t-R)].
\label{eq:bbgw-detector-dipole}
\end{equation}
Radial orbital motion of neutral matter produces a vector field with a nonzero tidal strain and a detector response for suitable arm orientations.

\begin{figure}[tbp]
\centering
\includegraphics[width=\columnwidth]{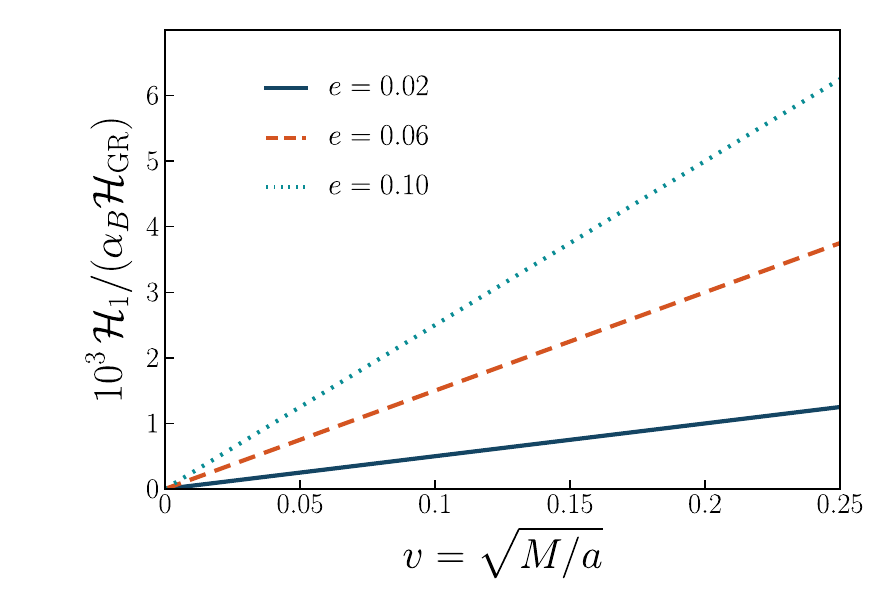}
\caption{Dipole strain scale relative to the circular GR scale, with the coupling removed: $10^3\mathcal{H}_1/(\alpha_B\mathcal{H}_{\mathrm{GR}})=10^3ev/4$. The eccentricities are $e=0.02$, $0.06$, and $0.10$ over $0\leq v\leq0.25$, with $v=0$ denoting the continuous weak field limit. Multiplying the ordinate by $10^{-3}\alpha_B$ restores the amplitude ratio before angular and detector projections.}
\label{fig:bb-strain-ratio}
\end{figure}

We compare the dipole amplitude with the circular general relativity scale $\mathcal H_{\mathrm{GR}}=4G\mu v^{2}/R$ and define
\begin{equation}
\begin{gathered}
\alpha_{B}=\frac{\xi^{2}b^{2}}{\kappa}=\frac{\ell^{2}}{\kappa b^{2}},\qquad \frac{\mathcal H_{1}}{\mathcal H_{\mathrm{GR}}}=\frac{\alpha_{B}}{4}ev.
\end{gathered}
\label{eq:bbgw-strain-comparison}
\end{equation}
The factor $ev$ originates in the radial angular source and distinguishes this dipole from radiation supplied by independent body charges. Figure~\ref{fig:bb-strain-ratio} removes $\alpha_B$ and shows the linear dependence on $v$ and $e$. The comparison concerns amplitude scales before angular and detector projections.

%%%%%%%%%%%%%%%%%%%%%%%%%%%%%%%%%%%%%%%%%%%%%%%%%%%%%%%%%%%%%%%%%%%%%%%%%%%%%%%%
\section{Energy carried by the axial dipole}
\label{sec:bbgw-flux}
%%%%%%%%%%%%%%%%%%%%%%%%%%%%%%%%%%%%%%%%%%%%%%%%%%%%%%%%%%%%%%%%%%%%%%%%%%%%%%%%

The leading vector energy is carried by $u_{10}$, whose amplitude is first order in the nonminimal coupling, while the metric strain is second order. We evaluate the luminosity from the Maxwell term of the action. Applying the general relativity tensor flux formula to $h_{\hat r\hat A}^{V}$ would assign an incorrect coupling order.

At leading order, in the large radius and small coupling expansions, a single real axial harmonic carries the outward flux
\begin{equation}
\begin{aligned}
P_{B,Lm}^{\infty}={}&-\lim_{R\rightarrow\infty}\int R^{2}\left\langle T^{F\,r}{}_{t}\right\rangle\mathrm{d}\Omega  = n_{L}\left\langle\left(\partial_{t}u_{Lm}^{\mathrm{out}}\right)^{2}\right\rangle.
\end{aligned}
\label{eq:bbgw-maxwell-flux}
\end{equation}
For an outgoing wave, $F_{tA}=\partial_{t}u\,S_{A}$ and $F_{rA}=-\partial_{t}u\,S_{A}+O(R^{-1})$. The angular integral gives $n_{L}$ through Eq.~\eqref{eq:bbgw-harmonic-normalization}. For $u_{10}^{\mathrm{out}}=\mathcal U_{1}\cos[\Omega(t-R)]$,
\begin{equation}
\begin{gathered}
\mathcal U_{1}=-\frac{\xi b\mu e C_{1}}{6}v^{3},\qquad n_{1}=2,\qquad \left\langle\left(\partial_{t}u_{10}^{\mathrm{out}}\right)^{2}\right\rangle=\frac{\Omega^{2}\mathcal U_{1}^{2}}{2}.
\end{gathered}
\label{eq:bbgw-dipole-flux-steps}
\end{equation}
The partial vector luminosity is
\begin{equation}
P_{B,10}^{\infty}=\frac{\xi^{2}b^{2}\mu^{2}}{48\pi M^{2}}e^{2}v^{12}+\cdots.
\label{eq:bbgw-final-dipole-flux}
\end{equation}
This result shows that the leading nonzero coupling and eccentricity orders. No outgoing general relativity tensor wave contributes to the angular dipole. The energy in the vector induced metric, its kinetic mixing with the vector, and corrections to the Maxwell normalization enter at higher coupling orders. Since $\delta X=0$, the potential supplies no quadratic contribution to this mode at the retained order.

\begin{figure}[tbp]
\centering
\includegraphics[width=\columnwidth]{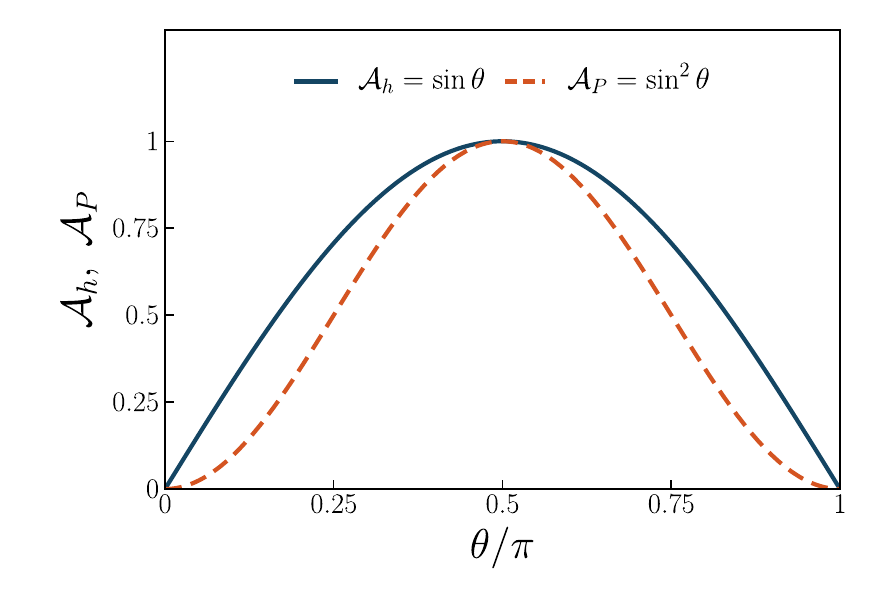}
\caption{Angular profiles of the leading axial dipole: the strain envelope $\mathcal A_h=\max_t|h_{\hat r\hat\phi}^{V,10}|/\mathcal{H}_1=\sin\theta$ and the scaled power density $\mathcal A_P=(8\pi/3)(P_{B,10}^{\infty})^{-1}\mathrm{d}P_{B,10}^{\infty}/\mathrm{d}\Omega=\sin^2\theta$, both with unit maximum. }
\label{fig:bb-angular-pattern}
\end{figure}

The dipole strain envelope is $\max_t|h_{\hat r\hat\phi}^{V,10}|/\mathcal{H}_1=\sin\theta$. Using $S_{\hat\phi}^{10}=-C_1\sin\theta$ and $\int\sin^2\theta\,\mathrm{d}\Omega=8\pi/3$, the Maxwell flux gives
\begin{equation}
\frac{1}{P_{B,10}^{\infty}}\frac{\mathrm{d}P_{B,10}^{\infty}}{\mathrm{d}\Omega}=\frac{3}{8\pi}\sin^2\theta.
\label{eq:bb-dipole-angular-power}
\end{equation}
In Fig.~\ref{fig:bb-angular-pattern}, both profiles vanish on the orbital axis and peak in the orbital plane. The power has the squared angular dependence of the strain, while a detector also applies the arm projection $F_V$.

The leading circular quadrupole luminosity in general relativity provides the reference~\cite{PetersMathews1963,Peters1964}
\begin{equation}
\begin{gathered}
P_{\mathrm{GR}}^{\mathrm{circ}}=\frac{32}{5}\frac{G\mu^{2}}{M^{2}}v^{10},\qquad \frac{P_{B,10}^{\infty}}{P_{\mathrm{GR}}^{\mathrm{circ}}}=\frac{5}{192}\alpha_{B}e^{2}v^{2}+\cdots \,.
\end{gathered}
\label{eq:bbgw-flux-comparison}
\end{equation}
The eccentric correction to the general relativity denominator enters beyond the retained order because the numerator is already quadratic in $e$. The suppression by $e^{2}v^{2}$ differs from dipole emission due to unequal independent scalar or vector charges, which can enter at a lower velocity order~\cite{Eardley1975,Foster2007,Maselli2020}. Here, the velocity dependence follows from the induced current in Eq.~\eqref{eq:bbgw-current-dust}.

\begin{figure}[tbp]
\centering
\includegraphics[width=\columnwidth]{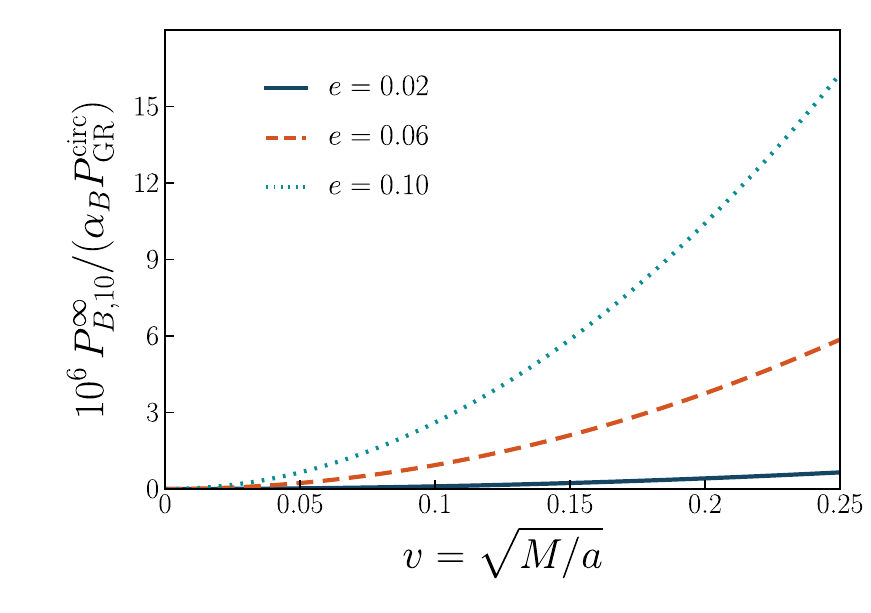}
\caption{Partial axial dipole luminosity relative to the circular GR quadrupole luminosity, with the coupling removed: $10^6P_{B,10}^{\infty}/(\alpha_BP_{\mathrm{GR}}^{\mathrm{circ}})=10^6(5/192)e^2v^2$. The eccentricities are $e=0.02$, $0.06$, and $0.10$, with $0\leq v\leq0.25$. }
\label{fig:bb-luminosity-ratio}
\end{figure}

Figure~\ref{fig:bb-luminosity-ratio} removes $\alpha_B$ and shows the relative scaling $e^2v^2$, which follows from $P_{B,10}^{\infty}\propto e^2v^{12}$ and $P_{\mathrm{GR}}^{\mathrm{circ}}\propto v^{10}$.

A complete inspiral calculation would combine the orbital energy and angular momentum with the tensor and vector fluxes at infinity and the flux through the inner boundary. The energy balance would take the form
\begin{equation}
\begin{gathered}
\frac{\mathrm{d}E_{\mathrm{orb}}}{\mathrm{d}t}=-P_{g}^{\infty}-P_{B}^{\infty}-P_{\mathrm{inner}},
\end{gathered}
\label{eq:bbgw-balance}
\end{equation}
where $P_{B}^{\infty}=P_{B,10}^{\infty}+P_{B,\mathrm{other}}^{\infty}$. Equation~\eqref{eq:bbgw-final-dipole-flux} supplies one partial contribution. The polar response and corrections are also needed for the complete phase evolution and bounds from inspiral observations.

%%%%%%%%%%%%%%%%%%%%%%%%%%%%%%%%%%%%%%%%%%%%%%%%%%%%%%%%%%%%%%%%%%%%%%%%%%%%%%%%
\section{Relation to the homogeneous spectrum and to previous waveforms}
\label{sec:bbgw-literature}
%%%%%%%%%%%%%%%%%%%%%%%%%%%%%%%%%%%%%%%%%%%%%%%%%%%%%%%%%%%%%%%%%%%%%%%%%%%%%%%%

%%%%%%%%%%%%%%%%%%%%%%%%%%%%%%%%%%%%%%%%%%%%%%%%%%%%%%%%%%%%%%%%%%%%%%%%%%%%%%%%
\subsection{Excitation amplitudes and gravitational master frequencies}
\label{subsec:bbgw-spectrum-response}
%%%%%%%%%%%%%%%%%%%%%%%%%%%%%%%%%%%%%%%%%%%%%%%%%%%%%%%%%%%%%%%%%%%%%%%%%%%%%%%%

The vector response contributes to the metric alongside the gravitational spectrum obtained in Ref.~\cite{Liu2026Decoupling}. In the observer region, the axial metric takes the schematic form
\begin{equation}
\begin{gathered}
h_{\mu\nu}^{\mathrm{axial}}=\mathcal R^{g}_{\mu\nu}\!\left[\Psi_{g}\right]+\mathcal R^{B}_{\mu\nu}\!\left[u\right],\qquad u=\widetilde G_{B}S_{B}.
\end{gathered}
\label{eq:bbgw-response-structure}
\end{equation}
The two terms describe the gravitational reconstruction and the vector response. The source integral and the poles of $\widetilde G_{B}$ affect the metric through $\mathcal R^{B}_{\mu\nu}$. In other words, the homogeneous gravitational frequencies leave the vector excitation undetermined~\cite{Berti2026Spectroscopy}. Equations~\eqref{eq:bbgw-dipole-leading-z}, \eqref{eq:bbgw-final-dipole-strain}, and \eqref{eq:bbgw-final-dipole-flux} give its amplitude, tidal strain, and partial luminosity for the specified orbit.

In charged black hole perturbation theory, coupled gravitational and electromagnetic equations allow an uncharged infalling particle to excite an electromagnetic response~\cite{Zerilli1974,Moncrief1974}. Here, the curvature induced vector source is $-\xi b_{\mu}T_{\mathrm{p}}^{\mu\nu}$, and its axial selection rule follows from $T_{rA}^{\mathrm{p}}$.

The perturbation equations depend on the background vector, as illustrated by stability analyses of charged bumblebee configurations~\cite{Mai2024} and studies of their coupled gravitoelectromagnetic spectra~\cite{Li2026ChargedQNMs}. Likewise, the polar reduction of Ref.~\cite{Liu2026Decoupling} assumes vacuum conditions on the norm and potential derivative. With matter present, the polar equations take into account the responses in Eqs.~\eqref{eq:bbgw-smooth-potential} and \eqref{eq:bbgw-radial-potential}. Furthermore, the scalar potential supplies no direct axial angular source at linear perturbation order.

%%%%%%%%%%%%%%%%%%%%%%%%%%%%%%%%%%%%%%%%%%%%%%%%%%%%%%%%%%%%%%%%%%%%%%%%%%%%%%%%%%%%%%%%%%%%%%%%%%%%%%%%%%%%%%%%%%%%%%%%%%%%%%%%%%%%%%%%%%%%%%%%%%%%%%%%%%%%%%%%%%%%%%%%%%%%%%%%%%%%%%%%%%%%%%%%%%%%%%%%%%%%%%%%%%%%%%%%%%%%%%%%%%%%%%%%%%%%%%%%%%%%%%%%%%%%%%%%%%%%%%%%%%%%%%%%%%%%%%%%%%%%%%%%%%%%%%%%%%%%%%%%%%%%%%%%%%%%%%%%%%%%%%%%%%%%%%%%%%%%%%%%%%%%%%%%%%%%%%%%%%%%%%%%%%%%%%%%%%%%%%%%%%%%%%%%%%%%%%%%%%%%%%%%%%%%%%%%%%%%%%%%%%%%%%%%%%%%%%%%%%%%%%%%%%%%%%%%%%%%%%%%%%%%%%%%%%%%%%%%%%%%%%%%%%%%%%%%%%%%%%%%%%%%%%%%%%%%%%%%%%%%%%%%%%%%%%%%%%%%%%%%%%%%%%%%%%%%%%%%%%%%%%%%%%%%%%%%%%%%%%%%%%%%%%%%%%%%%%%%%%%%%%%%%%%%%%%%%%%%%%%%%%%%%%%%%%%%%%%%%%%%%%%%%%%%%%%%%%%%%%%%%%%%%%%%%%%%%%%%%%%%%%%%%%%%%%%%%%%%%%%%%%%%%%%%%%%%%%%%%%%%%%%%%%%

\subsection{Physical clocks and propagation}
\label{subsec:bbgw-propagation}

The master equations express characteristic speeds in $(t,r_*)$ coordinates. Conversion to physical speeds uses the matter metric. For a static observer, $\mathrm{d}\tau_{\mathrm{loc}}=\sqrt f\,\mathrm{d}t$, $\mathrm{d}\mathcal R_{\mathrm{loc}}=\sqrt{q/f}\,\mathrm{d}r$, and $\frac{\mathrm{d}\mathcal R_{\mathrm{loc}}}{\mathrm{d}\tau_{\mathrm{loc}}}=\sqrt q\,\frac{\mathrm{d}r_*}{\mathrm{d}t}$. A minimally coupled radial photon has unit local speed; the tensor, vector, and photon characteristics in this frame are $v_{T}^{2}=q$, $v_{B}^{2}=q/\gamma$, and $v_{\gamma}^{2}=1$, respectively. In this regard, the Schwarzschild form of the tensor operator retains a dependence on the conversion to matter clocks and local distances. The outgoing vector phase is $\omega(t-c_{\mathrm{eff}}^{-1}r_*)$, whose zero coupling and weak field limit enters the explicit strains. When propagation accumulates a substantial coupling dependent phase, the full phase must be kept.

%%%%%%%%%%%%%%%%%%%%%%%%%%%%%%%%%%%%%%%%%%%%%%%%%%%%%%%%%%%%%%%%%%%%%%%%%%%%%%%%%%%%%%%%%%%%%%%%%%%%%%%%%%%%%%%%%%%%%%%%%%%%%%%%%%%%%%%%%%%%%%%%%%%%%%%%%%%%%%%%%%%%%%%%%%%%%%%%%%%%%%%%%%%%%%%%%%%%%%%%%%%%%%%%%%%%%%%%%%%%%%%%%%%%%%%%%%%%%%%%%%%%%%%%%%%%%%%%%%%%%%%%%%%%%%%%%%%%%%%%%%%%%%%%%%%%%%%%%%%%%%%%%%%%%%%%%%%%%%%%%%%%%%%%%%%%%%%%%%%%%%%%%%%%%%%%%%%%%%%%%%%%%%%%%%%%%%%%%%%%%%%%%%%%%%%%%%%%%%%%%%%%%%%%%%%%%%%%%%%%%%%%%%%%%%%%%%%%%%%%%%%%%%%%%%%%%%%%%%%%%%%%%%%%%%%%%%%%%%%%%%%%%%%%%%%%%%%%%%%%%%%%%%%%%%%%%%%%%%%%%%%%%%%%%%%%%%%%%%%%%%%%%%%%%%%%%%%%%%%%%%%%%%%%%%%%%%
\subsection{From orbital waveforms to field generated waveforms}
\label{subsec:bbgw-waveform-comparison}
%%%%%%%%%%%%%%%%%%%%%%%%%%%%%%%%%%%%%%%%%%%%%%%%%%%%%%%%%%%%%%%%%%%%%%%%%%%%%%%%

Shi et al.~\cite{Shi2026PeriodicOrbits} evaluated the standard quadrupole formula on periodic geodesics in bumblebee black hole geometries. Long et al.~\cite{Long2026EMRI} constructed inspiral waveforms within the Augmented Analytic Kludge approximation. These descriptions incorporate the geometry through the orbital motion and its quadrupole moment.

Here, we derive the axial vector source from the field equation and obtain its outgoing amplitudes and tidal strain through the retarded response and metric reconstruction. This calculation adds a vector polarization and its partial luminosity to the radiation associated with the modified orbit. The explicit coefficients describe weak fields, small eccentricity, and the retained coupling order. In contrast, forecasts based on an orbital quadrupole model apply to that model's waveform assumptions; Eq.~\eqref{eq:bbgw-final-dipole-strain} requires its own inspiral and detector analysis.

The weak field analyses in Refs.~\cite{Amarilo2024,AraujoFilho2026MetricAffine,AraujoFilho2026SME,AraujoFilho2026Propagation} derived retarded waveforms from localized sources. In the present exterior, the radial Green function takes into account curvature scattering, while the reconstructed tidal tensor identifies the polarization. The source coefficients follow from the chosen action and vector background, whose orientation also affects the polarizations~\cite{Liang2022}.

%%%%%%%%%%%%%%%%%%%%%%%%%%%%%%%%%%%%%%%%%%%%%%%%%%%%%%%%%%%%%%%%%%%%%%%%%%%%%%%%%%%%%%%%%%%%%%%%%%%%%%%%%%%%%%%%%%%%%%%%%%%%%%%%%%%%%%%%%%%%%%%%%%%%%%%%%%%%%%%%%%%%%%%%%%%%%%%%%%%%%%%%%%%%%%%%%%%%%%%%%%%%%%%%%%%%%%%%%%%%%%%%%%%%%%%%%%%%%%%%%%%%%%%%%%%%%%%%%%%%%%%%%%%%%%%%%%%%%%%%%%%%%%%%%%%%%%%%%%%%%%%%%%%%%%%%%%%%%%%%%%%%%%%%%%%%%%%%%%%%%%%%%%%%%%%%%%%%%%%%%%%%%%%%%%%%%%%%%%%%%%%%%%%%%%%%%%%%%%%%%%%%%%%%%%%%%%%%%%%%%%%%%%%%%%%%%%%%%%%%%%%%%%%%%%%%%%%%%%%%%%%%%%%%%%%%%%%%%%%%%%%%%%%%%%%%%%%%%%%%%%%%

\section{Concluding remarks}
\label{sec:bbgw-conclusions}

We derived the leading axial bumblebee radiation generated by a neutral point particle around a black hole with a radial vector vacuum. The nonminimal curvature coupling converted the matter stress into an effective vector current, and current conservation fixed the localized circular potential response to $\delta V'=\xi\rho/4$ in the absence of an independent radial boundary current. Radial motion instead introduced a conservation equation with boundary data.

The axial angular projection selected the radial transport of angular momentum. Its source vanished for circular motion, whereas a small eccentricity excited the $(L,m)=(1,0)$ dipole at $\Omega$ and the calculated $(2,1)$ contribution at $2\Omega$. Their outgoing coefficients scaled as $ev^3$ and $ev^4$. We reconstructed the angular dipole through its invariant equations and obtained a nonzero radial transverse tidal strain proportional to $\xi^2b^2\mu ev^3\sin\theta/R$. The signal decreased as the inverse observer radius, vanished on the orbital axis, and reached its largest amplitude in the orbital plane. The corresponding arm projection was nonzero for suitable detector orientations.

The Maxwell stress yielded a partial dipole luminosity proportional to $\xi^2b^2\mu^2e^2v^{12}/M^2$, with angular power proportional to $\sin^2\theta$. The vector amplitude was first order in the coupling, whereas both its Maxwell luminosity and the induced metric strain were second order. Relative to the circular general relativity scales, the strain amplitude ratio was $\alpha_Bev/4$ and the luminosity ratio was $5\alpha_Be^2v^2/192$, with $\alpha_B=\xi^2b^2/\kappa$. In this regard, the induced dipole exhibited an additional suppression by eccentricity and orbital velocity. The source integral and metric reconstruction supplied radiative information beyond the homogeneous gravitational master frequencies.

The explicit coefficients described the leading weak field and small eccentricity terms at the maintained coupling orders. The radial Green function specified an exterior problem with an ingoing master boundary condition; the background vector required an additional inner description for a global extension. The analysis also identified the sourced polar response, the effective action of a compact secondary, and the remaining energy fluxes as the additional ingredients for an inspiral waveform.

\appendix

%%%%%%%%%%%%%%%%%%%%%%%%%%%%%%%%%%%%%%%%%%%%%%%%%%%%%%%%%%%%%%%%%%%%%%%%%%%%%%%%
\section{Details of the tidal reconstruction}
\label{app:bbgw-curvature}
%%%%%%%%%%%%%%%%%%%%%%%%%%%%%%%%%%%%%%%%%%%%%%%%%%%%%%%%%%%%%%%%%%%%%%%%%%%%%%%%

For a metric with zero shift and lapse $N=\sqrt f$, write $\mathrm{d}s^{2}=-N^{2}\mathrm{d}t^{2}+\gamma_{ij}\mathrm{d}x^{i}\mathrm{d}x^{j}$. At linear order about a static spatial metric, the terms quadratic in $\partial_t\gamma_{ij}$ vanish, giving
\begin{equation}
R_{titj}=N D_{i}D_{j}N-\frac12\partial_{t}^{2}\gamma_{ij}+O(h^{2}).
\label{eq:bbgw-adm-curvature}
\end{equation}
Here, $D_i$ is the spatial derivative associated with $\gamma_{ij}$. For $\delta\gamma_{rA}=H S_A$, the connection variations are
\begin{equation}
\begin{gathered}
\delta\Gamma^{r}_{rA}=-\frac{fH}{qr}S_{A},\qquad \delta\Gamma^{r}_{AB}=\frac{fH}{q}S_{AB}.
\end{gathered}
\label{eq:bbgw-spatial-connections}
\end{equation}
Using $NN'=f'/2$ gives Eq.~\eqref{eq:bbgw-coordinate-curvature}. To first order in the vector metric perturbation, an orthonormal frame for a static observer is
\begin{equation}
\begin{gathered}
e_{\hat0}=f^{-1/2}\partial_t,\quad e_{\hat r}=\sqrt{\frac fq}\partial_r,\quad e_{\hat A}=\frac{\mathsf e_{\hat A}{}^{B}}{r}\partial_B-\frac{fH}{qr}S_{\hat A}\partial_r.
\end{gathered}
\label{eq:bbgw-observer-frame}
\end{equation}
The radial angular tide contains both the coordinate curvature variation and the radial correction to $e_{\hat A}$. With $\bar R_{trtr}=f''/2$, it becomes
\begin{equation}
\begin{aligned}
\delta\mathcal E_{\hat r\hat A}={}&\Biggl[-\frac{\partial_t^{2}H}{2r\sqrt{qf}}+\frac{ff'H}{2qr^{2}\sqrt{qf}} -\frac{f''H}{2qr}\sqrt{\frac fq}\Biggr]S_{\hat A}.
\end{aligned}
\label{eq:bbgw-frame-curvature-step}
\end{equation}
The derivatives of $f$ and the relation between $H$ and $u$ combine the last two terms into $3\ell M u/(bq^{2}r^{4})$, reproducing Eq.~\eqref{eq:bbgw-orthonormal-curvature}. A small rotation of the spatial triad changes the background tidal terms. The radiative $1/R$ contribution is unchanged, since those terms decay as $M/R^{4}$.

%%%%%%%%%%%%%%%%%%%%%%%%%%%%%%%%%%%%%%%%%%%%%%%%%%%%%%%%%%%%%%%%%%%%%%%%%%%%%%%%
\section{A derivative sign in the axial metric equation}
\label{app:bbgw-sign}
%%%%%%%%%%%%%%%%%%%%%%%%%%%%%%%%%%%%%%%%%%%%%%%%%%%%%%%%%%%%%%%%%%%%%%%%%%%%%%%%

The time angular metric equation provides a direct check of the dipole reconstruction. With the harmonic conventions of Eq.~\eqref{eq:bbgw-harmonic-definitions}, the Schwarzschild curvature variation is
\begin{equation}
\begin{aligned}
\frac{\delta R_{tA}}{S_A}={}&\frac f2\left(\partial_t\partial_r h_1-\partial_r^{2}h_0\right) +\frac fr\partial_t h_1 +\left(\frac{n_L}{2r^{2}}-\frac{f'}r\right)h_0.
\end{aligned}
\label{eq:bbgw-gr-time-angular}
\end{equation}
In the normalization of Ref.~\cite{Liu2026Decoupling}, this curvature variation and the vector reconstruction give
\begin{equation}
\begin{aligned}
0={}&\partial_r^{2}h_0-\partial_t\partial_r h_1-\frac2r\partial_t h_1 -\frac{D_L+2f}{r^{2}f}h_0\\ &+\frac{\ell}{fb_r}\left[\partial_t\partial_r u-\frac{1-5f}{2rf}\partial_tu\right].
\end{aligned}
\label{eq:bbgw-corrected-time-angular}
\end{equation}
The mixed derivative $\partial_t\partial_r h_1$ carries a minus sign, whereas Eq.~(4.1) of that reference prints a plus sign. With $a_u=\ell/(fb_r)$, $h_0=0$, and $h_1=a_u u$, the vector reconstruction in its Eqs.~(5.1) and (5.2) satisfies Eq.~\eqref{eq:bbgw-corrected-time-angular}, since $a_u'/a_u=-f'/(2f)$. The printed sign instead leaves
\begin{equation}
a_u\left[2\partial_t\partial_r u-\frac{f'}f\partial_tu\right].
\label{eq:bbgw-printed-sign-residual}
\end{equation}
The same sign follows from the general relativity contribution in Eq.~\eqref{eq:bbgw-gr-time-angular}. We use Eq.~\eqref{eq:bbgw-corrected-time-angular} in the main text and check the dipole through its invariant equation. The source projection follows directly from the covariant vector equation.

%%%%%%%%%%%%%%%%%%%%%%%%%%%%%%%%%%%%%%%%%%%%%%%%%%%%%%%%%%%%%%%%%%%%%%%%%%%%%%%%
\section{Exterior vector profile and the inner boundary}
\label{app:bbgw-exterior}
%%%%%%%%%%%%%%%%%%%%%%%%%%%%%%%%%%%%%%%%%%%%%%%%%%%%%%%%%%%%%%%%%%%%%%%%%%%%%%%%

The metric in Eq.~\eqref{eq:bbgw-background} admits the ingoing coordinate $v_{\mathrm{in}}=t+\sqrt q\,r_*$. In this chart,
\begin{equation}
\begin{aligned}
\mathrm{d}\bar s^{2}={}&-f\,\mathrm{d}v_{\mathrm{in}}^{2}+2\sqrt q\,\mathrm{d}v_{\mathrm{in}}\mathrm{d}r +r^{2}\mathrm{d}\Omega^{2},\qquad \bar B={}&b\sqrt{\frac qf}\,\mathrm{d}r.
\end{aligned}
\label{eq:bbgw-ingoing-background}
\end{equation}
The metric extends through $f=0$ in this chart, while the radial vector component diverges. Vector regularity also involves derivatives, as illustrated by the scalar
\begin{equation}
\begin{gathered}
\left(\bar\nabla_{\mu}\bar B_{\nu}\right)\left(\bar\nabla^{\mu}\bar B^{\nu}\right) \frac{b^{2}}q\left[\frac{(f')^{2}}{4f}+\frac{2f}{r^{2}}\right].
\end{gathered}
\label{eq:bbgw-vector-derivative-invariant}
\end{equation}
The first term diverges as $r\rightarrow2M$ from the exterior. The finite master functions in Eq.~\eqref{eq:bbgw-static-matching} describe the radial wave equation. The Green function used here imposes an ingoing master boundary condition on this exterior problem. Furthermore, regular vector continuation has instead been constructed in geometries with additional supporting matter~\cite{AraujoFilho2026SimpsonVisser}, which involve a different background.

%%%%%%%%%%%%%%%%%%%%%%%%%%%%%%%%%%%%%%%%%%%%%%%%%%%%%%%%%%%%%%%%%%%%%%%%%%%%%%%%%%%%%%%%%%%%%%%%%%%%%%%%%%%%%%%%%%%%%%%%%%%%%%%%%%%%%%%%%%%%%%%%%%%%%%%%%%%%%%%%%%%%%%%%%%%%%%%%%%%%%%%%%%%%%%%%%%%%%%%%%%%%%%%%%%%%%%%%%%%%%%%%%%%%%%%%%%%%%%%%%%%%%%%%%%%%%%%%%%%%%%%%%%%%%%%%%%%%%%%%%%%%%%%%%%%%%%%%%%%%%%%%%%%%%%%%%%%%%%%%%%%%%%%%%%%%%%%%%%%%%%%%%%%%%%%%%%%%%%%%%%%%%%%%%%%%%%%%%%%%%%%%%%%%%%%%%%%%%%%%%%%%%%%%%%%%%%%%%%%%%%%%%%%%%%%%%%%%%%%%%%%%%%%%%%%%%%%%%%%%%%%%%%%%%%%%%%%%%%%%%%%%%%%%%%%%%%%%%%%%%%%%%%

\section*{Acknowledgments}
\hspace{0.5cm} A. A. Araújo Filho is supported by Conselho Nacional de Desenvolvimento Cient\'{\i}fico e Tecnol\'{o}gico (CNPq) -- [150223/2025-0].

%%%%%%%%%%%%%%%%%%%%%%%%%%%%%%%%%%%%%%%%%%%%%%%%%%%%%%%%%%%%%%%%%%%%%%%%%%%%%%%%%%%%%%%%%%
\section*{Data Availability Statement}

Data associated with this study consist of the analytical expressions and numerical figures presented in the manuscript. No additional dataset is required to reproduce the analytical results. The code used to generate the numerical results and figures is available from the corresponding author upon reasonable request.

\bibliography{main}

%merlin.mbs apsrev4-1.bst 2010-07-25 4.21a (PWD, AO, DPC) hacked
%Control: key (0)
%Control: author (0) dotless jnrlst
%Control: editor formatted (1) identically to author
%Control: production of article title (0) allowed
%Control: page (1) range
%Control: year (0) verbatim
%Control: production of eprint (0) enabled
\begin{thebibliography}{56}%
\makeatletter
\providecommand \@ifxundefined [1]{%
 \@ifx{#1\undefined}
}%
\providecommand \@ifnum [1]{%
 \ifnum #1\expandafter \@firstoftwo
 \else \expandafter \@secondoftwo
 \fi
}%
\providecommand \@ifx [1]{%
 \ifx #1\expandafter \@firstoftwo
 \else \expandafter \@secondoftwo
 \fi
}%
\providecommand \natexlab [1]{#1}%
\providecommand \enquote  [1]{``#1''}%
\providecommand \bibnamefont  [1]{#1}%
\providecommand \bibfnamefont [1]{#1}%
\providecommand \citenamefont [1]{#1}%
\providecommand \href@noop [0]{\@secondoftwo}%
\providecommand \href [0]{\begingroup \@sanitize@url \@href}%
\providecommand \@href[1]{\@@startlink{#1}\@@href}%
\providecommand \@@href[1]{\endgroup#1\@@endlink}%
\providecommand \@sanitize@url [0]{\catcode `\\12\catcode `\$12\catcode
  `\&12\catcode `\#12\catcode `\^12\catcode `\_12\catcode `\%12\relax}%
\providecommand \@@startlink[1]{}%
\providecommand \@@endlink[0]{}%
\providecommand \url  [0]{\begingroup\@sanitize@url \@url }%
\providecommand \@url [1]{\endgroup\@href {#1}{\urlprefix }}%
\providecommand \urlprefix  [0]{URL }%
\providecommand \Eprint [0]{\href }%
\providecommand \doibase [0]{http://dx.doi.org/}%
\providecommand \selectlanguage [0]{\@gobble}%
\providecommand \bibinfo  [0]{\@secondoftwo}%
\providecommand \bibfield  [0]{\@secondoftwo}%
\providecommand \translation [1]{[#1]}%
\providecommand \BibitemOpen [0]{}%
\providecommand \bibitemStop [0]{}%
\providecommand \bibitemNoStop [0]{.\EOS\space}%
\providecommand \EOS [0]{\spacefactor3000\relax}%
\providecommand \BibitemShut  [1]{\csname bibitem#1\endcsname}%
\let\auto@bib@innerbib\@empty
%</preamble>
\bibitem [{\citenamefont {Abbott}\ \emph {et~al.}(2016)\citenamefont {Abbott}
  \emph {et~al.}}]{Abbott2016Observation}%
  \BibitemOpen
  \bibfield  {author} {\bibinfo {author} {\bibfnamefont {B.~P.}\ \bibnamefont
  {Abbott}} \emph {et~al.} (\bibinfo {collaboration} {LIGO Scientific and
  Virgo}),\ }\bibfield  {title} {\enquote {\bibinfo {title} {Observation of
  gravitational waves from a binary black hole merger},}\ }\href {\doibase
  10.1103/PhysRevLett.116.061102} {\bibfield  {journal} {\bibinfo  {journal}
  {Phys. Rev. Lett.}\ }\textbf {\bibinfo {volume} {116}},\ \bibinfo {pages}
  {061102} (\bibinfo {year} {2016})},\ \Eprint
  {http://arxiv.org/abs/1602.03837} {arXiv:1602.03837 [gr-qc]} \BibitemShut
  {NoStop}%
\bibitem [{\citenamefont {Abbott}\ \emph {et~al.}(2017)\citenamefont {Abbott}
  \emph {et~al.}}]{Abbott2017Speed}%
  \BibitemOpen
  \bibfield  {author} {\bibinfo {author} {\bibfnamefont {B.~P.}\ \bibnamefont
  {Abbott}} \emph {et~al.} (\bibinfo {collaboration} {LIGO Scientific, Virgo,
  Fermi-GBM and INTEGRAL}),\ }\bibfield  {title} {\enquote {\bibinfo {title}
  {Gravitational waves and gamma-rays from a binary neutron star merger:
  {GW170817} and {GRB 170817A}},}\ }\href {\doibase 10.3847/2041-8213/aa920c}
  {\bibfield  {journal} {\bibinfo  {journal} {Astrophys. J. Lett.}\ }\textbf
  {\bibinfo {volume} {848}},\ \bibinfo {pages} {L13} (\bibinfo {year}
  {2017})},\ \Eprint {http://arxiv.org/abs/1710.05834} {arXiv:1710.05834
  [astro-ph.HE]} \BibitemShut {NoStop}%
\bibitem [{\citenamefont {Kosteleck{\'y}}\ and\ \citenamefont
  {Mewes}(2016)}]{KosteleckyMewes2016}%
  \BibitemOpen
  \bibfield  {author} {\bibinfo {author} {\bibfnamefont {V.~Alan}\ \bibnamefont
  {Kosteleck{\'y}}}\ and\ \bibinfo {author} {\bibfnamefont {Matthew}\
  \bibnamefont {Mewes}},\ }\bibfield  {title} {\enquote {\bibinfo {title}
  {Testing local {Lorentz} invariance with gravitational waves},}\ }\href
  {\doibase 10.1016/j.physletb.2016.04.040} {\bibfield  {journal} {\bibinfo
  {journal} {Phys. Lett. B}\ }\textbf {\bibinfo {volume} {757}},\ \bibinfo
  {pages} {510--514} (\bibinfo {year} {2016})},\ \Eprint
  {http://arxiv.org/abs/1602.04782} {arXiv:1602.04782 [gr-qc]} \BibitemShut
  {NoStop}%
\bibitem [{\citenamefont {Mewes}(2019)}]{Mewes2019}%
  \BibitemOpen
  \bibfield  {author} {\bibinfo {author} {\bibfnamefont {Matthew}\ \bibnamefont
  {Mewes}},\ }\bibfield  {title} {\enquote {\bibinfo {title} {Signals for
  {Lorentz} violation in gravitational waves},}\ }\href {\doibase
  10.1103/PhysRevD.99.104062} {\bibfield  {journal} {\bibinfo  {journal} {Phys.
  Rev. D}\ }\textbf {\bibinfo {volume} {99}},\ \bibinfo {pages} {104062}
  (\bibinfo {year} {2019})},\ \Eprint {http://arxiv.org/abs/1905.00409}
  {arXiv:1905.00409 [gr-qc]} \BibitemShut {NoStop}%
\bibitem [{\citenamefont {Kosteleck{\'y}}\ and\ \citenamefont
  {Samuel}(1989)}]{KosteleckySamuel1989}%
  \BibitemOpen
  \bibfield  {author} {\bibinfo {author} {\bibfnamefont {V.~Alan}\ \bibnamefont
  {Kosteleck{\'y}}}\ and\ \bibinfo {author} {\bibfnamefont {Stuart}\
  \bibnamefont {Samuel}},\ }\bibfield  {title} {\enquote {\bibinfo {title}
  {Spontaneous breaking of {Lorentz} symmetry in string theory},}\ }\href
  {\doibase 10.1103/PhysRevD.39.683} {\bibfield  {journal} {\bibinfo  {journal}
  {Phys. Rev. D}\ }\textbf {\bibinfo {volume} {39}},\ \bibinfo {pages} {683}
  (\bibinfo {year} {1989})}\BibitemShut {NoStop}%
\bibitem [{\citenamefont {Kosteleck{\'y}}(2004)}]{Kostelecky2004}%
  \BibitemOpen
  \bibfield  {author} {\bibinfo {author} {\bibfnamefont {V.~Alan}\ \bibnamefont
  {Kosteleck{\'y}}},\ }\bibfield  {title} {\enquote {\bibinfo {title} {Gravity,
  {Lorentz} violation, and the standard model},}\ }\href {\doibase
  10.1103/PhysRevD.69.105009} {\bibfield  {journal} {\bibinfo  {journal} {Phys.
  Rev. D}\ }\textbf {\bibinfo {volume} {69}},\ \bibinfo {pages} {105009}
  (\bibinfo {year} {2004})},\ \Eprint {http://arxiv.org/abs/hep-th/0312310}
  {arXiv:hep-th/0312310} \BibitemShut {NoStop}%
\bibitem [{\citenamefont {Bluhm}\ and\ \citenamefont
  {Kosteleck{\'y}}(2005)}]{BluhmKostelecky2005}%
  \BibitemOpen
  \bibfield  {author} {\bibinfo {author} {\bibfnamefont {Robert}\ \bibnamefont
  {Bluhm}}\ and\ \bibinfo {author} {\bibfnamefont {V.~Alan}\ \bibnamefont
  {Kosteleck{\'y}}},\ }\bibfield  {title} {\enquote {\bibinfo {title}
  {Spontaneous {Lorentz} violation, {Nambu-Goldstone} modes, and gravity},}\
  }\href {\doibase 10.1103/PhysRevD.71.065008} {\bibfield  {journal} {\bibinfo
  {journal} {Phys. Rev. D}\ }\textbf {\bibinfo {volume} {71}},\ \bibinfo
  {pages} {065008} (\bibinfo {year} {2005})},\ \Eprint
  {http://arxiv.org/abs/hep-th/0412320} {arXiv:hep-th/0412320} \BibitemShut
  {NoStop}%
\bibitem [{\citenamefont {Ara{\'u}jo~Filho}\ \emph
  {et~al.}(2026{\natexlab{a}})\citenamefont {Ara{\'u}jo~Filho}, \citenamefont
  {Heidari}, \citenamefont {Lobo},\ and\ \citenamefont
  {Bezerra}}]{AraujoFilho:2025zaj}%
  \BibitemOpen
  \bibfield  {author} {\bibinfo {author} {\bibfnamefont {A.~A.}\ \bibnamefont
  {Ara{\'u}jo~Filho}}, \bibinfo {author} {\bibfnamefont {N.}~\bibnamefont
  {Heidari}}, \bibinfo {author} {\bibfnamefont {Iarley~P.}\ \bibnamefont
  {Lobo}}, \ and\ \bibinfo {author} {\bibfnamefont {V.~B.}\ \bibnamefont
  {Bezerra}},\ }\bibfield  {title} {\enquote {\bibinfo {title} {{Gravitational
  aspects of a new bumblebee black hole}},}\ }\href {\doibase
  10.1016/j.aop.2026.170469} {\bibfield  {journal} {\bibinfo  {journal} {Annals
  Phys.}\ }\textbf {\bibinfo {volume} {489}},\ \bibinfo {pages} {170469}
  (\bibinfo {year} {2026}{\natexlab{a}})},\ \Eprint
  {http://arxiv.org/abs/2511.12839} {arXiv:2511.12839 [gr-qc]} \BibitemShut
  {NoStop}%
\bibitem [{\citenamefont {Ara{\'u}jo~Filho}(2025)}]{AraujoFilho:2025hkm}%
  \BibitemOpen
  \bibfield  {author} {\bibinfo {author} {\bibfnamefont {A.~A.}\ \bibnamefont
  {Ara{\'u}jo~Filho}},\ }\bibfield  {title} {\enquote {\bibinfo {title} {{How
  does non-metricity affect particle creation and evaporation in bumblebee
  gravity?}}}\ }\href {\doibase 10.1088/1475-7516/2025/06/026} {\bibfield
  {journal} {\bibinfo  {journal} {JCAP}\ }\textbf {\bibinfo {volume} {06}},\
  \bibinfo {pages} {026} (\bibinfo {year} {2025})},\ \bibinfo {note} {[Erratum:
  JCAP 02, E01 (2026)]},\ \Eprint {http://arxiv.org/abs/2501.00927}
  {arXiv:2501.00927 [gr-qc]} \BibitemShut {NoStop}%
\bibitem [{\citenamefont {Bluhm}\ \emph {et~al.}(2008)\citenamefont {Bluhm},
  \citenamefont {Fung},\ and\ \citenamefont {Kosteleck{\'y}}}]{Bluhm2008}%
  \BibitemOpen
  \bibfield  {author} {\bibinfo {author} {\bibfnamefont {Robert}\ \bibnamefont
  {Bluhm}}, \bibinfo {author} {\bibfnamefont {Shu-Hong}\ \bibnamefont {Fung}},
  \ and\ \bibinfo {author} {\bibfnamefont {V.~Alan}\ \bibnamefont
  {Kosteleck{\'y}}},\ }\bibfield  {title} {\enquote {\bibinfo {title}
  {Spontaneous {Lorentz} and diffeomorphism violation, massive modes, and
  gravity},}\ }\href {\doibase 10.1103/PhysRevD.77.065020} {\bibfield
  {journal} {\bibinfo  {journal} {Phys. Rev. D}\ }\textbf {\bibinfo {volume}
  {77}},\ \bibinfo {pages} {065020} (\bibinfo {year} {2008})},\ \Eprint
  {http://arxiv.org/abs/0712.4119} {arXiv:0712.4119 [hep-th]} \BibitemShut
  {NoStop}%
\bibitem [{\citenamefont {Liang}\ \emph {et~al.}(2022)\citenamefont {Liang},
  \citenamefont {Xu}, \citenamefont {Lu},\ and\ \citenamefont
  {Shao}}]{Liang2022}%
  \BibitemOpen
  \bibfield  {author} {\bibinfo {author} {\bibfnamefont {Dicong}\ \bibnamefont
  {Liang}}, \bibinfo {author} {\bibfnamefont {Rui}\ \bibnamefont {Xu}},
  \bibinfo {author} {\bibfnamefont {Xuchen}\ \bibnamefont {Lu}}, \ and\
  \bibinfo {author} {\bibfnamefont {Lijing}\ \bibnamefont {Shao}},\ }\bibfield
  {title} {\enquote {\bibinfo {title} {Polarizations of gravitational waves in
  the bumblebee gravity model},}\ }\href {\doibase 10.1103/PhysRevD.106.124019}
  {\bibfield  {journal} {\bibinfo  {journal} {Phys. Rev. D}\ }\textbf {\bibinfo
  {volume} {106}},\ \bibinfo {pages} {124019} (\bibinfo {year} {2022})},\
  \Eprint {http://arxiv.org/abs/2207.14423} {arXiv:2207.14423 [gr-qc]}
  \BibitemShut {NoStop}%
\bibitem [{\citenamefont {Amarilo}\ \emph {et~al.}(2024)\citenamefont
  {Amarilo}, \citenamefont {Ferreira~Filho}, \citenamefont {Ara{\'u}jo~Filho},\
  and\ \citenamefont {Reis}}]{Amarilo2024}%
  \BibitemOpen
  \bibfield  {author} {\bibinfo {author} {\bibfnamefont {K.~M.}\ \bibnamefont
  {Amarilo}}, \bibinfo {author} {\bibfnamefont {M.~B.}\ \bibnamefont
  {Ferreira~Filho}}, \bibinfo {author} {\bibfnamefont {A.~A.}\ \bibnamefont
  {Ara{\'u}jo~Filho}}, \ and\ \bibinfo {author} {\bibfnamefont {J.~A. A.~S.}\
  \bibnamefont {Reis}},\ }\bibfield  {title} {\enquote {\bibinfo {title}
  {Gravitational waves effects in a {Lorentz}-violating scenario},}\ }\href
  {\doibase 10.1016/j.physletb.2024.138785} {\bibfield  {journal} {\bibinfo
  {journal} {Phys. Lett. B}\ }\textbf {\bibinfo {volume} {855}},\ \bibinfo
  {pages} {138785} (\bibinfo {year} {2024})},\ \Eprint
  {http://arxiv.org/abs/2307.10937} {arXiv:2307.10937 [gr-qc]} \BibitemShut
  {NoStop}%
\bibitem [{\citenamefont
  {Ara{\'u}jo~Filho}(2026{\natexlab{a}})}]{AraujoFilho2026MetricAffine}%
  \BibitemOpen
  \bibfield  {author} {\bibinfo {author} {\bibfnamefont {A.~A.}\ \bibnamefont
  {Ara{\'u}jo~Filho}},\ }\bibfield  {title} {\enquote {\bibinfo {title}
  {Gravitational waves in metric-affine bumblebee gravity},}\ }\href {\doibase
  10.1140/epjc/s10052-026-15814-w} {\bibfield  {journal} {\bibinfo  {journal}
  {Eur. Phys. J. C}\ }\textbf {\bibinfo {volume} {86}},\ \bibinfo {pages} {578}
  (\bibinfo {year} {2026}{\natexlab{a}})},\ \Eprint
  {http://arxiv.org/abs/2603.08310} {arXiv:2603.08310 [gr-qc]} \BibitemShut
  {NoStop}%
\bibitem [{\citenamefont {Ara{\'u}jo~Filho}\ \emph
  {et~al.}(2026{\natexlab{b}})\citenamefont {Ara{\'u}jo~Filho}, \citenamefont
  {Heidari},\ and\ \citenamefont {Lobo}}]{AraujoFilho2026SME}%
  \BibitemOpen
  \bibfield  {author} {\bibinfo {author} {\bibfnamefont {A.~A.}\ \bibnamefont
  {Ara{\'u}jo~Filho}}, \bibinfo {author} {\bibfnamefont {N.}~\bibnamefont
  {Heidari}}, \ and\ \bibinfo {author} {\bibfnamefont {Iarley~P.}\ \bibnamefont
  {Lobo}},\ }\bibfield  {title} {\enquote {\bibinfo {title} {Gravitational
  waves in a minimal gravitational {SME}},}\ }\href {\doibase
  10.1016/j.physletb.2026.140350} {\bibfield  {journal} {\bibinfo  {journal}
  {Phys. Lett. B}\ }\textbf {\bibinfo {volume} {875}},\ \bibinfo {pages}
  {140350} (\bibinfo {year} {2026}{\natexlab{b}})},\ \Eprint
  {http://arxiv.org/abs/2602.09183} {arXiv:2602.09183 [gr-qc]} \BibitemShut
  {NoStop}%
\bibitem [{\citenamefont {Ara{\'u}jo~Filho}\ \emph
  {et~al.}(2026{\natexlab{c}})\citenamefont {Ara{\'u}jo~Filho}, \citenamefont
  {Heidari},\ and\ \citenamefont {Lobo}}]{AraujoFilho2026Propagation}%
  \BibitemOpen
  \bibfield  {author} {\bibinfo {author} {\bibfnamefont {A.~A.}\ \bibnamefont
  {Ara{\'u}jo~Filho}}, \bibinfo {author} {\bibfnamefont {N.}~\bibnamefont
  {Heidari}}, \ and\ \bibinfo {author} {\bibfnamefont {Iarley~P.}\ \bibnamefont
  {Lobo}},\ }\bibfield  {title} {\enquote {\bibinfo {title} {Propagation
  effects of {Lorentz} violation in gravitational waves},}\ }\href {\doibase
  10.1140/epjc/s10052-026-15859-x} {\bibfield  {journal} {\bibinfo  {journal}
  {Eur. Phys. J. C}\ }\textbf {\bibinfo {volume} {86}},\ \bibinfo {pages} {630}
  (\bibinfo {year} {2026}{\natexlab{c}})},\ \Eprint
  {http://arxiv.org/abs/2602.19186} {arXiv:2602.19186 [gr-qc]} \BibitemShut
  {NoStop}%
\bibitem [{\citenamefont {Casana}\ \emph {et~al.}(2018)\citenamefont {Casana},
  \citenamefont {Cavalcante}, \citenamefont {Poulis},\ and\ \citenamefont
  {Santos}}]{Casana2018}%
  \BibitemOpen
  \bibfield  {author} {\bibinfo {author} {\bibfnamefont {R.}~\bibnamefont
  {Casana}}, \bibinfo {author} {\bibfnamefont {A.}~\bibnamefont {Cavalcante}},
  \bibinfo {author} {\bibfnamefont {F.~P.}\ \bibnamefont {Poulis}}, \ and\
  \bibinfo {author} {\bibfnamefont {E.~B.}\ \bibnamefont {Santos}},\ }\bibfield
   {title} {\enquote {\bibinfo {title} {Exact {Schwarzschild}-like solution in
  a bumblebee gravity model},}\ }\href {\doibase 10.1103/PhysRevD.97.104001}
  {\bibfield  {journal} {\bibinfo  {journal} {Phys. Rev. D}\ }\textbf {\bibinfo
  {volume} {97}},\ \bibinfo {pages} {104001} (\bibinfo {year} {2018})},\
  \Eprint {http://arxiv.org/abs/1711.02273} {arXiv:1711.02273 [gr-qc]}
  \BibitemShut {NoStop}%
\bibitem [{\citenamefont {Maluf}\ and\ \citenamefont
  {Neves}(2021)}]{MalufNeves2021}%
  \BibitemOpen
  \bibfield  {author} {\bibinfo {author} {\bibfnamefont {R.~V.}\ \bibnamefont
  {Maluf}}\ and\ \bibinfo {author} {\bibfnamefont {Juliano C.~S.}\ \bibnamefont
  {Neves}},\ }\bibfield  {title} {\enquote {\bibinfo {title} {Black holes with
  a cosmological constant in bumblebee gravity},}\ }\href {\doibase
  10.1103/PhysRevD.103.044002} {\bibfield  {journal} {\bibinfo  {journal}
  {Phys. Rev. D}\ }\textbf {\bibinfo {volume} {103}},\ \bibinfo {pages}
  {044002} (\bibinfo {year} {2021})},\ \Eprint
  {http://arxiv.org/abs/2011.12841} {arXiv:2011.12841 [gr-qc]} \BibitemShut
  {NoStop}%
\bibitem [{\citenamefont {Ding}\ and\ \citenamefont
  {Chen}(2021)}]{DingChen2021}%
  \BibitemOpen
  \bibfield  {author} {\bibinfo {author} {\bibfnamefont {Chikun}\ \bibnamefont
  {Ding}}\ and\ \bibinfo {author} {\bibfnamefont {Xiongwen}\ \bibnamefont
  {Chen}},\ }\bibfield  {title} {\enquote {\bibinfo {title} {Slowly rotating
  {Einstein}-bumblebee black hole solution and its greybody factor in a
  {Lorentz} violation model},}\ }\href {\doibase 10.1088/1674-1137/abce51}
  {\bibfield  {journal} {\bibinfo  {journal} {Chin. Phys. C}\ }\textbf
  {\bibinfo {volume} {45}},\ \bibinfo {pages} {025106} (\bibinfo {year}
  {2021})},\ \Eprint {http://arxiv.org/abs/2008.10474} {arXiv:2008.10474
  [gr-qc]} \BibitemShut {NoStop}%
\bibitem [{\citenamefont {Poulis}\ and\ \citenamefont
  {Soares}(2022)}]{PoulisSoares2022}%
  \BibitemOpen
  \bibfield  {author} {\bibinfo {author} {\bibfnamefont {F.~P.}\ \bibnamefont
  {Poulis}}\ and\ \bibinfo {author} {\bibfnamefont {M.~A.~C.}\ \bibnamefont
  {Soares}},\ }\bibfield  {title} {\enquote {\bibinfo {title} {Exact
  modifications on a vacuum spacetime due to a gradient bumblebee field at its
  vacuum expectation value},}\ }\href {\doibase
  10.1140/epjc/s10052-022-10547-y} {\bibfield  {journal} {\bibinfo  {journal}
  {Eur. Phys. J. C}\ }\textbf {\bibinfo {volume} {82}},\ \bibinfo {pages} {613}
  (\bibinfo {year} {2022})},\ \Eprint {http://arxiv.org/abs/2112.04040}
  {arXiv:2112.04040 [gr-qc]} \BibitemShut {NoStop}%
\bibitem [{\citenamefont {Xu}\ \emph {et~al.}(2023)\citenamefont {Xu},
  \citenamefont {Liang},\ and\ \citenamefont {Shao}}]{XuLiangShao2023}%
  \BibitemOpen
  \bibfield  {author} {\bibinfo {author} {\bibfnamefont {Rui}\ \bibnamefont
  {Xu}}, \bibinfo {author} {\bibfnamefont {Dicong}\ \bibnamefont {Liang}}, \
  and\ \bibinfo {author} {\bibfnamefont {Lijing}\ \bibnamefont {Shao}},\
  }\bibfield  {title} {\enquote {\bibinfo {title} {Static spherical vacuum
  solutions in the bumblebee gravity model},}\ }\href {\doibase
  10.1103/PhysRevD.107.024011} {\bibfield  {journal} {\bibinfo  {journal}
  {Phys. Rev. D}\ }\textbf {\bibinfo {volume} {107}},\ \bibinfo {pages}
  {024011} (\bibinfo {year} {2023})},\ \Eprint
  {http://arxiv.org/abs/2209.02209} {arXiv:2209.02209 [gr-qc]} \BibitemShut
  {NoStop}%
\bibitem [{\citenamefont {Ara{\'u}jo~Filho}\ \emph
  {et~al.}(2026{\natexlab{d}})\citenamefont {Ara{\'u}jo~Filho}, \citenamefont
  {Heidari}, \citenamefont {Lobo}, \citenamefont {Shi},\ and\ \citenamefont
  {Lobo}}]{AraujoFilho:2025rvn}%
  \BibitemOpen
  \bibfield  {author} {\bibinfo {author} {\bibfnamefont {A.~A.}\ \bibnamefont
  {Ara{\'u}jo~Filho}}, \bibinfo {author} {\bibfnamefont {N.}~\bibnamefont
  {Heidari}}, \bibinfo {author} {\bibfnamefont {Iarley~P.}\ \bibnamefont
  {Lobo}}, \bibinfo {author} {\bibfnamefont {Yuxuan}\ \bibnamefont {Shi}}, \
  and\ \bibinfo {author} {\bibfnamefont {Francisco S.~N.}\ \bibnamefont
  {Lobo}},\ }\bibfield  {title} {\enquote {\bibinfo {title} {{The flight of the
  bumblebee in a non-commutative geometry: A new black hole solution}},}\
  }\href {\doibase 10.1016/j.aop.2026.170487} {\bibfield  {journal} {\bibinfo
  {journal} {Annals Phys.}\ }\textbf {\bibinfo {volume} {490}},\ \bibinfo
  {pages} {170487} (\bibinfo {year} {2026}{\natexlab{d}})},\ \Eprint
  {http://arxiv.org/abs/2509.17867} {arXiv:2509.17867 [gr-qc]} \BibitemShut
  {NoStop}%
\bibitem [{\citenamefont {Liu}\ \emph {et~al.}(2025)\citenamefont {Liu},
  \citenamefont {Guo}, \citenamefont {Wei},\ and\ \citenamefont
  {Liu}}]{Liu2025Charged}%
  \BibitemOpen
  \bibfield  {author} {\bibinfo {author} {\bibfnamefont {Jia-Zhou}\
  \bibnamefont {Liu}}, \bibinfo {author} {\bibfnamefont {Wen-Di}\ \bibnamefont
  {Guo}}, \bibinfo {author} {\bibfnamefont {Shao-Wen}\ \bibnamefont {Wei}}, \
  and\ \bibinfo {author} {\bibfnamefont {Yu-Xiao}\ \bibnamefont {Liu}},\
  }\bibfield  {title} {\enquote {\bibinfo {title} {Charged spherically
  symmetric and slowly rotating charged black hole solutions in bumblebee
  gravity},}\ }\href {\doibase 10.1140/epjc/s10052-025-13859-x} {\bibfield
  {journal} {\bibinfo  {journal} {Eur. Phys. J. C}\ }\textbf {\bibinfo {volume}
  {85}},\ \bibinfo {pages} {145} (\bibinfo {year} {2025})},\ \Eprint
  {http://arxiv.org/abs/2407.08396} {arXiv:2407.08396 [gr-qc]} \BibitemShut
  {NoStop}%
\bibitem [{\citenamefont {Zhu}\ and\ \citenamefont {Li}(2026)}]{ZhuLi2026}%
  \BibitemOpen
  \bibfield  {author} {\bibinfo {author} {\bibfnamefont {Jie}\ \bibnamefont
  {Zhu}}\ and\ \bibinfo {author} {\bibfnamefont {Hao}\ \bibnamefont {Li}},\
  }\bibfield  {title} {\enquote {\bibinfo {title} {Full classification of
  static spherical vacuum solutions to bumblebee gravity with general
  {VEVs}},}\ }\href {\doibase 10.1016/j.physletb.2026.140396} {\bibfield
  {journal} {\bibinfo  {journal} {Phys. Lett. B}\ }\textbf {\bibinfo {volume}
  {876}},\ \bibinfo {pages} {140396} (\bibinfo {year} {2026})},\ \Eprint
  {http://arxiv.org/abs/2511.03231} {arXiv:2511.03231 [gr-qc]} \BibitemShut
  {NoStop}%
\bibitem [{\citenamefont {Liu}\ \emph {et~al.}(2026{\natexlab{a}})\citenamefont
  {Liu}, \citenamefont {Wu}, \citenamefont {Wei},\ and\ \citenamefont
  {Liu}}]{Liu2026Vacua}%
  \BibitemOpen
  \bibfield  {author} {\bibinfo {author} {\bibfnamefont {Jia-Zhou}\
  \bibnamefont {Liu}}, \bibinfo {author} {\bibfnamefont {Shan-Ping}\
  \bibnamefont {Wu}}, \bibinfo {author} {\bibfnamefont {Shao-Wen}\ \bibnamefont
  {Wei}}, \ and\ \bibinfo {author} {\bibfnamefont {Yu-Xiao}\ \bibnamefont
  {Liu}},\ }\bibfield  {title} {\enquote {\bibinfo {title} {Exact black hole
  solutions in bumblebee gravity with lightlike or spacelike {VEVs}},}\ }\href
  {\doibase 10.1007/s11433-026-2961-8} {\bibfield  {journal} {\bibinfo
  {journal} {Sci. China Phys. Mech. Astron.}\ }\textbf {\bibinfo {volume}
  {69}},\ \bibinfo {pages} {270411} (\bibinfo {year} {2026}{\natexlab{a}})},\
  \Eprint {http://arxiv.org/abs/2510.16731} {arXiv:2510.16731 [gr-qc]}
  \BibitemShut {NoStop}%
\bibitem [{\citenamefont {Ara{\'u}jo~Filho}\ \emph {et~al.}(2023)\citenamefont
  {Ara{\'u}jo~Filho}, \citenamefont {Nascimento}, \citenamefont {Petrov},\ and\
  \citenamefont {Porf{\'i}rio}}]{AraujoFilho2023Vacuum}%
  \BibitemOpen
  \bibfield  {author} {\bibinfo {author} {\bibfnamefont {A.~A.}\ \bibnamefont
  {Ara{\'u}jo~Filho}}, \bibinfo {author} {\bibfnamefont {J.~R.}\ \bibnamefont
  {Nascimento}}, \bibinfo {author} {\bibfnamefont {A.~Yu.}\ \bibnamefont
  {Petrov}}, \ and\ \bibinfo {author} {\bibfnamefont {P.~J.}\ \bibnamefont
  {Porf{\'i}rio}},\ }\bibfield  {title} {\enquote {\bibinfo {title} {Vacuum
  solution within a metric-affine bumblebee gravity},}\ }\href {\doibase
  10.1103/PhysRevD.108.085010} {\bibfield  {journal} {\bibinfo  {journal}
  {Phys. Rev. D}\ }\textbf {\bibinfo {volume} {108}},\ \bibinfo {pages}
  {085010} (\bibinfo {year} {2023})},\ \Eprint
  {http://arxiv.org/abs/2211.11821} {arXiv:2211.11821 [gr-qc]} \BibitemShut
  {NoStop}%
\bibitem [{\citenamefont {Ara{\'u}jo~Filho}\ \emph
  {et~al.}(2024{\natexlab{a}})\citenamefont {Ara{\'u}jo~Filho}, \citenamefont
  {Nascimento}, \citenamefont {Petrov},\ and\ \citenamefont
  {Porf{\'i}rio}}]{AraujoFilho2024Axisymmetric}%
  \BibitemOpen
  \bibfield  {author} {\bibinfo {author} {\bibfnamefont {A.~A.}\ \bibnamefont
  {Ara{\'u}jo~Filho}}, \bibinfo {author} {\bibfnamefont {J.~R.}\ \bibnamefont
  {Nascimento}}, \bibinfo {author} {\bibfnamefont {A.~Yu.}\ \bibnamefont
  {Petrov}}, \ and\ \bibinfo {author} {\bibfnamefont {P.~J.}\ \bibnamefont
  {Porf{\'i}rio}},\ }\bibfield  {title} {\enquote {\bibinfo {title} {An exact
  stationary axisymmetric vacuum solution within a metric-affine bumblebee
  gravity},}\ }\href {\doibase 10.1088/1475-7516/2024/07/004} {\bibfield
  {journal} {\bibinfo  {journal} {JCAP}\ }\textbf {\bibinfo {volume} {07}},\
  \bibinfo {pages} {004} (\bibinfo {year} {2024}{\natexlab{a}})},\ \Eprint
  {http://arxiv.org/abs/2402.13014} {arXiv:2402.13014 [gr-qc]} \BibitemShut
  {NoStop}%
\bibitem [{\citenamefont {Ara{\'u}jo~Filho}\ \emph
  {et~al.}(2026{\natexlab{e}})\citenamefont {Ara{\'u}jo~Filho}, \citenamefont
  {Heidari}, \citenamefont {Bezerra},\ and\ \citenamefont
  {Lobo}}]{AraujoFilho2026GeneralStatic}%
  \BibitemOpen
  \bibfield  {author} {\bibinfo {author} {\bibfnamefont {A.~A.}\ \bibnamefont
  {Ara{\'u}jo~Filho}}, \bibinfo {author} {\bibfnamefont {N.}~\bibnamefont
  {Heidari}}, \bibinfo {author} {\bibfnamefont {V.~B.}\ \bibnamefont
  {Bezerra}}, \ and\ \bibinfo {author} {\bibfnamefont {Francisco S.~N.}\
  \bibnamefont {Lobo}},\ }\bibfield  {title} {\enquote {\bibinfo {title} {A
  general static and spherically symmetric black hole in traceless
  metric-affine bumblebee gravity},}\ }\href {https://arxiv.org/abs/2609.29675}
  {\bibfield  {journal} {\bibinfo  {journal} {arXiv e-prints}\ } (\bibinfo
  {year} {2026}{\natexlab{e}})},\ \bibinfo {note} {preprint},\ \Eprint
  {http://arxiv.org/abs/2609.29675} {arXiv:2609.29675 [gr-qc]} \BibitemShut
  {NoStop}%
\bibitem [{\citenamefont
  {Ara{\'u}jo~Filho}(2026{\natexlab{b}})}]{AraujoFilho2026SimpsonVisser}%
  \BibitemOpen
  \bibfield  {author} {\bibinfo {author} {\bibfnamefont {A.~A.}\ \bibnamefont
  {Ara{\'u}jo~Filho}},\ }\bibfield  {title} {\enquote {\bibinfo {title} {An
  exact {Simpson-Visser} family in bumblebee gravity},}\ }\href
  {https://arxiv.org/abs/2610.21701} {\bibfield  {journal} {\bibinfo  {journal}
  {arXiv e-prints}\ } (\bibinfo {year} {2026}{\natexlab{b}})},\ \bibinfo {note}
  {preprint},\ \Eprint {http://arxiv.org/abs/2610.21701} {arXiv:2610.21701
  [gr-qc]} \BibitemShut {NoStop}%
\bibitem [{\citenamefont {Oliveira}\ \emph {et~al.}(2021)\citenamefont
  {Oliveira}, \citenamefont {Dantas},\ and\ \citenamefont
  {Almeida}}]{Oliveira2021QNMs}%
  \BibitemOpen
  \bibfield  {author} {\bibinfo {author} {\bibfnamefont {R.}~\bibnamefont
  {Oliveira}}, \bibinfo {author} {\bibfnamefont {D.~M.}\ \bibnamefont
  {Dantas}}, \ and\ \bibinfo {author} {\bibfnamefont {C.~A.~S.}\ \bibnamefont
  {Almeida}},\ }\bibfield  {title} {\enquote {\bibinfo {title} {Quasinormal
  frequencies for a black hole in a bumblebee gravity},}\ }\href {\doibase
  10.1209/0295-5075/ac130c} {\bibfield  {journal} {\bibinfo  {journal} {EPL}\
  }\textbf {\bibinfo {volume} {135}},\ \bibinfo {pages} {10003} (\bibinfo
  {year} {2021})},\ \Eprint {http://arxiv.org/abs/2105.07956} {arXiv:2105.07956
  [gr-qc]} \BibitemShut {NoStop}%
\bibitem [{\citenamefont {Ara{\'u}jo~Filho}\ \emph
  {et~al.}(2024{\natexlab{b}})\citenamefont {Ara{\'u}jo~Filho}, \citenamefont
  {Hassanabadi}, \citenamefont {Heidari}, \citenamefont {K{\v r}{\'i}{\v z}},\
  and\ \citenamefont {Zare}}]{AraujoFilho2024Traces}%
  \BibitemOpen
  \bibfield  {author} {\bibinfo {author} {\bibfnamefont {A.~A.}\ \bibnamefont
  {Ara{\'u}jo~Filho}}, \bibinfo {author} {\bibfnamefont {H.}~\bibnamefont
  {Hassanabadi}}, \bibinfo {author} {\bibfnamefont {N.}~\bibnamefont
  {Heidari}}, \bibinfo {author} {\bibfnamefont {J.}~\bibnamefont {K{\v
  r}{\'i}{\v z}}}, \ and\ \bibinfo {author} {\bibfnamefont {S.}~\bibnamefont
  {Zare}},\ }\bibfield  {title} {\enquote {\bibinfo {title} {Gravitational
  traces of bumblebee gravity in metric-affine formalism},}\ }\href {\doibase
  10.1088/1361-6382/ad1712} {\bibfield  {journal} {\bibinfo  {journal} {Class.
  Quantum Grav.}\ }\textbf {\bibinfo {volume} {41}},\ \bibinfo {pages} {055003}
  (\bibinfo {year} {2024}{\natexlab{b}})},\ \Eprint
  {http://arxiv.org/abs/2305.18871} {arXiv:2305.18871 [gr-qc]} \BibitemShut
  {NoStop}%
\bibitem [{\citenamefont {Heidari}\ \emph {et~al.}(2024)\citenamefont
  {Heidari}, \citenamefont {Macedo}, \citenamefont {Ara{\'u}jo~Filho},\ and\
  \citenamefont {Hassanabadi}}]{Heidari2024Scattering}%
  \BibitemOpen
  \bibfield  {author} {\bibinfo {author} {\bibfnamefont {N.}~\bibnamefont
  {Heidari}}, \bibinfo {author} {\bibfnamefont {Caio F.~B.}\ \bibnamefont
  {Macedo}}, \bibinfo {author} {\bibfnamefont {A.~A.}\ \bibnamefont
  {Ara{\'u}jo~Filho}}, \ and\ \bibinfo {author} {\bibfnamefont
  {H.}~\bibnamefont {Hassanabadi}},\ }\bibfield  {title} {\enquote {\bibinfo
  {title} {Scattering effects of bumblebee gravity in metric-affine
  formalism},}\ }\href {\doibase 10.1140/epjc/s10052-024-13580-1} {\bibfield
  {journal} {\bibinfo  {journal} {Eur. Phys. J. C}\ }\textbf {\bibinfo {volume}
  {84}},\ \bibinfo {pages} {1221} (\bibinfo {year} {2024})},\ \Eprint
  {http://arxiv.org/abs/2407.05321} {arXiv:2407.05321 [gr-qc]} \BibitemShut
  {NoStop}%
\bibitem [{\citenamefont {Liu}\ \emph {et~al.}(2023)\citenamefont {Liu},
  \citenamefont {Fang}, \citenamefont {Jing},\ and\ \citenamefont
  {Wang}}]{Liu2023SlowQNMs}%
  \BibitemOpen
  \bibfield  {author} {\bibinfo {author} {\bibfnamefont {Wentao}\ \bibnamefont
  {Liu}}, \bibinfo {author} {\bibfnamefont {Xiongjun}\ \bibnamefont {Fang}},
  \bibinfo {author} {\bibfnamefont {Jiliang}\ \bibnamefont {Jing}}, \ and\
  \bibinfo {author} {\bibfnamefont {Jieci}\ \bibnamefont {Wang}},\ }\bibfield
  {title} {\enquote {\bibinfo {title} {{QNMs} of slowly rotating
  {Einstein}--bumblebee black hole},}\ }\href {\doibase
  10.1140/epjc/s10052-023-11231-5} {\bibfield  {journal} {\bibinfo  {journal}
  {Eur. Phys. J. C}\ }\textbf {\bibinfo {volume} {83}},\ \bibinfo {pages} {83}
  (\bibinfo {year} {2023})},\ \Eprint {http://arxiv.org/abs/2211.03156}
  {arXiv:2211.03156 [gr-qc]} \BibitemShut {NoStop}%
\bibitem [{\citenamefont {Mai}\ \emph {et~al.}(2024)\citenamefont {Mai},
  \citenamefont {Xu}, \citenamefont {Liang},\ and\ \citenamefont
  {Shao}}]{Mai2024}%
  \BibitemOpen
  \bibfield  {author} {\bibinfo {author} {\bibfnamefont {Zhan-Feng}\
  \bibnamefont {Mai}}, \bibinfo {author} {\bibfnamefont {Rui}\ \bibnamefont
  {Xu}}, \bibinfo {author} {\bibfnamefont {Dicong}\ \bibnamefont {Liang}}, \
  and\ \bibinfo {author} {\bibfnamefont {Lijing}\ \bibnamefont {Shao}},\
  }\bibfield  {title} {\enquote {\bibinfo {title} {Dynamic instability analysis
  for bumblebee black holes: The odd parity},}\ }\href {\doibase
  10.1103/PhysRevD.109.084076} {\bibfield  {journal} {\bibinfo  {journal}
  {Phys. Rev. D}\ }\textbf {\bibinfo {volume} {109}},\ \bibinfo {pages}
  {084076} (\bibinfo {year} {2024})},\ \Eprint
  {http://arxiv.org/abs/2401.07757} {arXiv:2401.07757 [gr-qc]} \BibitemShut
  {NoStop}%
\bibitem [{\citenamefont {Li}\ \emph {et~al.}(2026)\citenamefont {Li},
  \citenamefont {Liu}, \citenamefont {Guo},\ and\ \citenamefont
  {Liu}}]{Li2026ChargedQNMs}%
  \BibitemOpen
  \bibfield  {author} {\bibinfo {author} {\bibfnamefont {Bo-Rui}\ \bibnamefont
  {Li}}, \bibinfo {author} {\bibfnamefont {Jia-Zhou}\ \bibnamefont {Liu}},
  \bibinfo {author} {\bibfnamefont {Wen-Di}\ \bibnamefont {Guo}}, \ and\
  \bibinfo {author} {\bibfnamefont {Yu-Xiao}\ \bibnamefont {Liu}},\ }\bibfield
  {title} {\enquote {\bibinfo {title} {Quasinormal modes of a charged
  spherically symmetric black hole in bumblebee gravity},}\ }\href {\doibase
  10.1140/epjc/s10052-026-16043-x} {\bibfield  {journal} {\bibinfo  {journal}
  {Eur. Phys. J. C}\ }\textbf {\bibinfo {volume} {86}},\ \bibinfo {pages} {974}
  (\bibinfo {year} {2026})},\ \Eprint {http://arxiv.org/abs/2510.20503}
  {arXiv:2510.20503 [gr-qc]} \BibitemShut {NoStop}%
\bibitem [{\citenamefont {Liu}\ \emph {et~al.}(2026{\natexlab{b}})\citenamefont
  {Liu}, \citenamefont {Liu}, \citenamefont {Liu}, \citenamefont {Su},\ and\
  \citenamefont {Zeng}}]{Liu2026Decoupling}%
  \BibitemOpen
  \bibfield  {author} {\bibinfo {author} {\bibfnamefont {Hui-Fa}\ \bibnamefont
  {Liu}}, \bibinfo {author} {\bibfnamefont {Wentao}\ \bibnamefont {Liu}},
  \bibinfo {author} {\bibfnamefont {Yu-Xiao}\ \bibnamefont {Liu}}, \bibinfo
  {author} {\bibfnamefont {Qi}~\bibnamefont {Su}}, \ and\ \bibinfo {author}
  {\bibfnamefont {Ding-fang}\ \bibnamefont {Zeng}},\ }\bibfield  {title}
  {\enquote {\bibinfo {title} {Gravitational-bumblebee perturbations: exact
  decoupling and isospectrality},}\ }\href {\doibase 10.1007/JHEP08(2026)027}
  {\bibfield  {journal} {\bibinfo  {journal} {JHEP}\ }\textbf {\bibinfo
  {volume} {08}},\ \bibinfo {pages} {027} (\bibinfo {year}
  {2026}{\natexlab{b}})},\ \Eprint {http://arxiv.org/abs/2605.02820}
  {arXiv:2605.02820 [gr-qc]} \BibitemShut {NoStop}%
\bibitem [{\citenamefont {Regge}\ and\ \citenamefont
  {Wheeler}(1957)}]{ReggeWheeler1957}%
  \BibitemOpen
  \bibfield  {author} {\bibinfo {author} {\bibfnamefont {Tullio}\ \bibnamefont
  {Regge}}\ and\ \bibinfo {author} {\bibfnamefont {John~A.}\ \bibnamefont
  {Wheeler}},\ }\bibfield  {title} {\enquote {\bibinfo {title} {Stability of a
  {Schwarzschild} singularity},}\ }\href {\doibase 10.1103/PhysRev.108.1063}
  {\bibfield  {journal} {\bibinfo  {journal} {Phys. Rev.}\ }\textbf {\bibinfo
  {volume} {108}},\ \bibinfo {pages} {1063--1069} (\bibinfo {year}
  {1957})}\BibitemShut {NoStop}%
\bibitem [{\citenamefont {Zerilli}(1970)}]{Zerilli1970}%
  \BibitemOpen
  \bibfield  {author} {\bibinfo {author} {\bibfnamefont {Frank~J.}\
  \bibnamefont {Zerilli}},\ }\bibfield  {title} {\enquote {\bibinfo {title}
  {Effective potential for even-parity {Regge-Wheeler} gravitational
  perturbation equations},}\ }\href {\doibase 10.1103/PhysRevLett.24.737}
  {\bibfield  {journal} {\bibinfo  {journal} {Phys. Rev. Lett.}\ }\textbf
  {\bibinfo {volume} {24}},\ \bibinfo {pages} {737--738} (\bibinfo {year}
  {1970})}\BibitemShut {NoStop}%
\bibitem [{\citenamefont {Sasaki}\ and\ \citenamefont
  {Tagoshi}(2003)}]{SasakiTagoshi2003}%
  \BibitemOpen
  \bibfield  {author} {\bibinfo {author} {\bibfnamefont {Misao}\ \bibnamefont
  {Sasaki}}\ and\ \bibinfo {author} {\bibfnamefont {Hideyuki}\ \bibnamefont
  {Tagoshi}},\ }\bibfield  {title} {\enquote {\bibinfo {title} {Analytic black
  hole perturbation approach to gravitational radiation},}\ }\href {\doibase
  10.12942/lrr-2003-6} {\bibfield  {journal} {\bibinfo  {journal} {Living Rev.
  Relativ.}\ }\textbf {\bibinfo {volume} {6}},\ \bibinfo {pages} {6} (\bibinfo
  {year} {2003})},\ \Eprint {http://arxiv.org/abs/gr-qc/0306120}
  {arXiv:gr-qc/0306120} \BibitemShut {NoStop}%
\bibitem [{\citenamefont {Berti}\ \emph {et~al.}(2009)\citenamefont {Berti},
  \citenamefont {Cardoso},\ and\ \citenamefont
  {Starinets}}]{BertiCardosoStarinets2009}%
  \BibitemOpen
  \bibfield  {author} {\bibinfo {author} {\bibfnamefont {Emanuele}\
  \bibnamefont {Berti}}, \bibinfo {author} {\bibfnamefont {Vitor}\ \bibnamefont
  {Cardoso}}, \ and\ \bibinfo {author} {\bibfnamefont {Andrei~O.}\ \bibnamefont
  {Starinets}},\ }\bibfield  {title} {\enquote {\bibinfo {title} {Quasinormal
  modes of black holes and black branes},}\ }\href {\doibase
  10.1088/0264-9381/26/16/163001} {\bibfield  {journal} {\bibinfo  {journal}
  {Class. Quantum Grav.}\ }\textbf {\bibinfo {volume} {26}},\ \bibinfo {pages}
  {163001} (\bibinfo {year} {2009})},\ \Eprint {http://arxiv.org/abs/0905.2975}
  {arXiv:0905.2975 [gr-qc]} \BibitemShut {NoStop}%
\bibitem [{\citenamefont {Berti}\ \emph {et~al.}(2026)\citenamefont {Berti},
  \citenamefont {Cardoso}, \citenamefont {Carullo} \emph
  {et~al.}}]{Berti2026Spectroscopy}%
  \BibitemOpen
  \bibfield  {author} {\bibinfo {author} {\bibfnamefont {Emanuele}\
  \bibnamefont {Berti}}, \bibinfo {author} {\bibfnamefont {Vitor}\ \bibnamefont
  {Cardoso}}, \bibinfo {author} {\bibfnamefont {Gregorio}\ \bibnamefont
  {Carullo}},  \emph {et~al.},\ }\bibfield  {title} {\enquote {\bibinfo {title}
  {Black hole spectroscopy: from theory to experiment},}\ }\href {\doibase
  10.1088/1361-6382/ae59e2} {\bibfield  {journal} {\bibinfo  {journal} {Class.
  Quantum Grav.}\ }\textbf {\bibinfo {volume} {43}},\ \bibinfo {pages} {123001}
  (\bibinfo {year} {2026})},\ \Eprint {http://arxiv.org/abs/2505.23895}
  {arXiv:2505.23895 [gr-qc]} \BibitemShut {NoStop}%
\bibitem [{\citenamefont {Liang}\ \emph {et~al.}(2023)\citenamefont {Liang},
  \citenamefont {Xu}, \citenamefont {Mai},\ and\ \citenamefont
  {Shao}}]{Liang2023VectorHair}%
  \BibitemOpen
  \bibfield  {author} {\bibinfo {author} {\bibfnamefont {Dicong}\ \bibnamefont
  {Liang}}, \bibinfo {author} {\bibfnamefont {Rui}\ \bibnamefont {Xu}},
  \bibinfo {author} {\bibfnamefont {Zhan-Feng}\ \bibnamefont {Mai}}, \ and\
  \bibinfo {author} {\bibfnamefont {Lijing}\ \bibnamefont {Shao}},\ }\bibfield
  {title} {\enquote {\bibinfo {title} {Probing vector hair of black holes with
  extreme-mass-ratio inspirals},}\ }\href {\doibase
  10.1103/PhysRevD.107.044053} {\bibfield  {journal} {\bibinfo  {journal}
  {Phys. Rev. D}\ }\textbf {\bibinfo {volume} {107}},\ \bibinfo {pages}
  {044053} (\bibinfo {year} {2023})},\ \Eprint
  {http://arxiv.org/abs/2212.09346} {arXiv:2212.09346 [gr-qc]} \BibitemShut
  {NoStop}%
\bibitem [{\citenamefont {Shi}\ \emph {et~al.}(2026)\citenamefont {Shi},
  \citenamefont {Zhang},\ and\ \citenamefont {Liu}}]{Shi2026PeriodicOrbits}%
  \BibitemOpen
  \bibfield  {author} {\bibinfo {author} {\bibfnamefont {Zijian}\ \bibnamefont
  {Shi}}, \bibinfo {author} {\bibfnamefont {Xiangdong}\ \bibnamefont {Zhang}},
  \ and\ \bibinfo {author} {\bibfnamefont {Yunlong}\ \bibnamefont {Liu}},\
  }\bibfield  {title} {\enquote {\bibinfo {title} {Periodic orbits and
  gravitational waveforms of black holes in bumblebee gravity},}\ }\href
  {\doibase 10.1016/j.dark.2026.102349} {\bibfield  {journal} {\bibinfo
  {journal} {Phys. Dark Univ.}\ }\textbf {\bibinfo {volume} {52}},\ \bibinfo
  {pages} {102349} (\bibinfo {year} {2026})},\ \Eprint
  {http://arxiv.org/abs/2603.14413} {arXiv:2603.14413 [gr-qc]} \BibitemShut
  {NoStop}%
\bibitem [{\citenamefont {Long}\ \emph {et~al.}(2027)\citenamefont {Long},
  \citenamefont {Xia}, \citenamefont {Gong}, \citenamefont {Cao}, \citenamefont
  {Pan},\ and\ \citenamefont {Jing}}]{Long2026EMRI}%
  \BibitemOpen
  \bibfield  {author} {\bibinfo {author} {\bibfnamefont {Sheng}\ \bibnamefont
  {Long}}, \bibinfo {author} {\bibfnamefont {Zhong-wu}\ \bibnamefont {Xia}},
  \bibinfo {author} {\bibfnamefont {Huajie}\ \bibnamefont {Gong}}, \bibinfo
  {author} {\bibfnamefont {Zhoujian}\ \bibnamefont {Cao}}, \bibinfo {author}
  {\bibfnamefont {Qiyuan}\ \bibnamefont {Pan}}, \ and\ \bibinfo {author}
  {\bibfnamefont {Jiliang}\ \bibnamefont {Jing}},\ }\bibfield  {title}
  {\enquote {\bibinfo {title} {Constraining {Lorentz} symmetry breaking in
  bumblebee gravity with extreme mass-ratio inspirals},}\ }\href {\doibase
  10.1088/1674-1137/aea1f0} {\bibfield  {journal} {\bibinfo  {journal} {Chin.
  Phys. C}\ }\textbf {\bibinfo {volume} {51}},\ \bibinfo {pages} {015105}
  (\bibinfo {year} {2027})},\ \bibinfo {note} {published online in 2026;
  assigned to the 2027 issue},\ \Eprint {http://arxiv.org/abs/2605.05362}
  {arXiv:2605.05362 [gr-qc]} \BibitemShut {NoStop}%
\bibitem [{\citenamefont {Bailey}\ \emph {et~al.}(2025)\citenamefont {Bailey},
  \citenamefont {Murray},\ and\ \citenamefont
  {Walter-Cardona}}]{Bailey2025Potential}%
  \BibitemOpen
  \bibfield  {author} {\bibinfo {author} {\bibfnamefont {Quentin~G.}\
  \bibnamefont {Bailey}}, \bibinfo {author} {\bibfnamefont {Hailey~S.}\
  \bibnamefont {Murray}}, \ and\ \bibinfo {author} {\bibfnamefont {Dario~T.}\
  \bibnamefont {Walter-Cardona}},\ }\bibfield  {title} {\enquote {\bibinfo
  {title} {Bumblebee gravity: Spherically symmetric solutions away from the
  potential minimum},}\ }\href {\doibase 10.1103/nkbj-vbjf} {\bibfield
  {journal} {\bibinfo  {journal} {Phys. Rev. D}\ }\textbf {\bibinfo {volume}
  {112}},\ \bibinfo {pages} {024069} (\bibinfo {year} {2025})},\ \Eprint
  {http://arxiv.org/abs/2503.10998} {arXiv:2503.10998 [gr-qc]} \BibitemShut
  {NoStop}%
\bibitem [{\citenamefont {Martel}\ and\ \citenamefont
  {Poisson}(2005)}]{MartelPoisson2005}%
  \BibitemOpen
  \bibfield  {author} {\bibinfo {author} {\bibfnamefont {Karl}\ \bibnamefont
  {Martel}}\ and\ \bibinfo {author} {\bibfnamefont {Eric}\ \bibnamefont
  {Poisson}},\ }\bibfield  {title} {\enquote {\bibinfo {title} {Gravitational
  perturbations of the {Schwarzschild} spacetime: A practical covariant and
  gauge-invariant formalism},}\ }\href {\doibase 10.1103/PhysRevD.71.104003}
  {\bibfield  {journal} {\bibinfo  {journal} {Phys. Rev. D}\ }\textbf {\bibinfo
  {volume} {71}},\ \bibinfo {pages} {104003} (\bibinfo {year} {2005})},\
  \Eprint {http://arxiv.org/abs/gr-qc/0502028} {arXiv:gr-qc/0502028}
  \BibitemShut {NoStop}%
\bibitem [{\citenamefont {Nagar}\ and\ \citenamefont
  {Rezzolla}(2005)}]{NagarRezzolla2005}%
  \BibitemOpen
  \bibfield  {author} {\bibinfo {author} {\bibfnamefont {Alessandro}\
  \bibnamefont {Nagar}}\ and\ \bibinfo {author} {\bibfnamefont {Luciano}\
  \bibnamefont {Rezzolla}},\ }\bibfield  {title} {\enquote {\bibinfo {title}
  {Gauge-invariant non-spherical metric perturbations of {Schwarzschild}
  black-hole spacetimes},}\ }\href {\doibase 10.1088/0264-9381/22/16/R01}
  {\bibfield  {journal} {\bibinfo  {journal} {Class. Quantum Grav.}\ }\textbf
  {\bibinfo {volume} {22}},\ \bibinfo {pages} {R167} (\bibinfo {year}
  {2005})},\ \bibinfo {note} {erratum: Class. Quantum Grav. 23, 4297 (2006),
  doi:10.1088/0264-9381/23/12/C01},\ \Eprint
  {http://arxiv.org/abs/gr-qc/0502064} {arXiv:gr-qc/0502064} \BibitemShut
  {NoStop}%
\bibitem [{\citenamefont {Eardley}(1975)}]{Eardley1975}%
  \BibitemOpen
  \bibfield  {author} {\bibinfo {author} {\bibfnamefont {Douglas~M.}\
  \bibnamefont {Eardley}},\ }\bibfield  {title} {\enquote {\bibinfo {title}
  {Observable effects of a scalar gravitational field in a binary pulsar},}\
  }\href {\doibase 10.1086/181744} {\bibfield  {journal} {\bibinfo  {journal}
  {Astrophys. J.}\ }\textbf {\bibinfo {volume} {196}},\ \bibinfo {pages}
  {L59--L62} (\bibinfo {year} {1975})}\BibitemShut {NoStop}%
\bibitem [{\citenamefont {Foster}(2007)}]{Foster2007}%
  \BibitemOpen
  \bibfield  {author} {\bibinfo {author} {\bibfnamefont {Brendan~Z.}\
  \bibnamefont {Foster}},\ }\bibfield  {title} {\enquote {\bibinfo {title}
  {Strong field effects on binary systems in {Einstein-aether} theory},}\
  }\href {\doibase 10.1103/PhysRevD.76.084033} {\bibfield  {journal} {\bibinfo
  {journal} {Phys. Rev. D}\ }\textbf {\bibinfo {volume} {76}},\ \bibinfo
  {pages} {084033} (\bibinfo {year} {2007})},\ \Eprint
  {http://arxiv.org/abs/0706.0704} {arXiv:0706.0704 [gr-qc]} \BibitemShut
  {NoStop}%
\bibitem [{\citenamefont {Barack}\ and\ \citenamefont
  {Pound}(2019)}]{BarackPound2019}%
  \BibitemOpen
  \bibfield  {author} {\bibinfo {author} {\bibfnamefont {Leor}\ \bibnamefont
  {Barack}}\ and\ \bibinfo {author} {\bibfnamefont {Adam}\ \bibnamefont
  {Pound}},\ }\bibfield  {title} {\enquote {\bibinfo {title} {Self-force and
  radiation reaction in general relativity},}\ }\href {\doibase
  10.1088/1361-6633/aae552} {\bibfield  {journal} {\bibinfo  {journal} {Rep.
  Prog. Phys.}\ }\textbf {\bibinfo {volume} {82}},\ \bibinfo {pages} {016904}
  (\bibinfo {year} {2019})},\ \Eprint {http://arxiv.org/abs/1805.10385}
  {arXiv:1805.10385 [gr-qc]} \BibitemShut {NoStop}%
\bibitem [{\citenamefont {Li}\ and\ \citenamefont {L{\"u}}(2026)}]{LiLu2026}%
  \BibitemOpen
  \bibfield  {author} {\bibinfo {author} {\bibfnamefont {Zhi-Chao}\
  \bibnamefont {Li}}\ and\ \bibinfo {author} {\bibfnamefont {H.}~\bibnamefont
  {L{\"u}}},\ }\bibfield  {title} {\enquote {\bibinfo {title} {When bumblebee
  meets {NLED}: {Lorentz}-violating black holes and regular spacetimes},}\
  }\href {\doibase 10.1140/epjc/s10052-026-16219-5} {\bibfield  {journal}
  {\bibinfo  {journal} {Eur. Phys. J. C}\ }\textbf {\bibinfo {volume} {86}},\
  \bibinfo {pages} {959} (\bibinfo {year} {2026})}\BibitemShut {NoStop}%
\bibitem [{\citenamefont {Peters}\ and\ \citenamefont
  {Mathews}(1963)}]{PetersMathews1963}%
  \BibitemOpen
  \bibfield  {author} {\bibinfo {author} {\bibfnamefont {P.~C.}\ \bibnamefont
  {Peters}}\ and\ \bibinfo {author} {\bibfnamefont {J.}~\bibnamefont
  {Mathews}},\ }\bibfield  {title} {\enquote {\bibinfo {title} {Gravitational
  radiation from point masses in a {Keplerian} orbit},}\ }\href {\doibase
  10.1103/PhysRev.131.435} {\bibfield  {journal} {\bibinfo  {journal} {Phys.
  Rev.}\ }\textbf {\bibinfo {volume} {131}},\ \bibinfo {pages} {435--440}
  (\bibinfo {year} {1963})}\BibitemShut {NoStop}%
\bibitem [{\citenamefont {Peters}(1964)}]{Peters1964}%
  \BibitemOpen
  \bibfield  {author} {\bibinfo {author} {\bibfnamefont {P.~C.}\ \bibnamefont
  {Peters}},\ }\bibfield  {title} {\enquote {\bibinfo {title} {Gravitational
  radiation and the motion of two point masses},}\ }\href {\doibase
  10.1103/PhysRev.136.B1224} {\bibfield  {journal} {\bibinfo  {journal} {Phys.
  Rev.}\ }\textbf {\bibinfo {volume} {136}},\ \bibinfo {pages} {B1224--B1232}
  (\bibinfo {year} {1964})}\BibitemShut {NoStop}%
\bibitem [{\citenamefont {Eardley}\ \emph {et~al.}(1973)\citenamefont
  {Eardley}, \citenamefont {Lee}, \citenamefont {Lightman}, \citenamefont
  {Wagoner},\ and\ \citenamefont {Will}}]{Eardley1973}%
  \BibitemOpen
  \bibfield  {author} {\bibinfo {author} {\bibfnamefont {Douglas~M.}\
  \bibnamefont {Eardley}}, \bibinfo {author} {\bibfnamefont {David~L.}\
  \bibnamefont {Lee}}, \bibinfo {author} {\bibfnamefont {Alan~P.}\ \bibnamefont
  {Lightman}}, \bibinfo {author} {\bibfnamefont {Robert~V.}\ \bibnamefont
  {Wagoner}}, \ and\ \bibinfo {author} {\bibfnamefont {Clifford~M.}\
  \bibnamefont {Will}},\ }\bibfield  {title} {\enquote {\bibinfo {title}
  {Gravitational-wave observations as a tool for testing relativistic
  gravity},}\ }\href {\doibase 10.1103/PhysRevLett.30.884} {\bibfield
  {journal} {\bibinfo  {journal} {Phys. Rev. Lett.}\ }\textbf {\bibinfo
  {volume} {30}},\ \bibinfo {pages} {884--886} (\bibinfo {year}
  {1973})}\BibitemShut {NoStop}%
\bibitem [{\citenamefont {Maselli}\ \emph {et~al.}(2020)\citenamefont
  {Maselli}, \citenamefont {Franchini}, \citenamefont {Gualtieri},\ and\
  \citenamefont {Sotiriou}}]{Maselli2020}%
  \BibitemOpen
  \bibfield  {author} {\bibinfo {author} {\bibfnamefont {Andrea}\ \bibnamefont
  {Maselli}}, \bibinfo {author} {\bibfnamefont {Nicola}\ \bibnamefont
  {Franchini}}, \bibinfo {author} {\bibfnamefont {Leonardo}\ \bibnamefont
  {Gualtieri}}, \ and\ \bibinfo {author} {\bibfnamefont {Thomas~P.}\
  \bibnamefont {Sotiriou}},\ }\bibfield  {title} {\enquote {\bibinfo {title}
  {Detecting scalar fields with extreme mass ratio inspirals},}\ }\href
  {\doibase 10.1103/PhysRevLett.125.141101} {\bibfield  {journal} {\bibinfo
  {journal} {Phys. Rev. Lett.}\ }\textbf {\bibinfo {volume} {125}},\ \bibinfo
  {pages} {141101} (\bibinfo {year} {2020})},\ \Eprint
  {http://arxiv.org/abs/2004.11895} {arXiv:2004.11895 [gr-qc]} \BibitemShut
  {NoStop}%
\bibitem [{\citenamefont {Zerilli}(1974)}]{Zerilli1974}%
  \BibitemOpen
  \bibfield  {author} {\bibinfo {author} {\bibfnamefont {Frank~J.}\
  \bibnamefont {Zerilli}},\ }\bibfield  {title} {\enquote {\bibinfo {title}
  {Perturbation analysis for gravitational and electromagnetic radiation in a
  {Reissner-Nordstr{\"o}m} geometry},}\ }\href {\doibase
  10.1103/PhysRevD.9.860} {\bibfield  {journal} {\bibinfo  {journal} {Phys.
  Rev. D}\ }\textbf {\bibinfo {volume} {9}},\ \bibinfo {pages} {860--868}
  (\bibinfo {year} {1974})}\BibitemShut {NoStop}%
\bibitem [{\citenamefont {Moncrief}(1974)}]{Moncrief1974}%
  \BibitemOpen
  \bibfield  {author} {\bibinfo {author} {\bibfnamefont {Vincent}\ \bibnamefont
  {Moncrief}},\ }\bibfield  {title} {\enquote {\bibinfo {title} {Stability of
  {Reissner-Nordstr{\"o}m} black holes},}\ }\href {\doibase
  10.1103/PhysRevD.10.1057} {\bibfield  {journal} {\bibinfo  {journal} {Phys.
  Rev. D}\ }\textbf {\bibinfo {volume} {10}},\ \bibinfo {pages} {1057}
  (\bibinfo {year} {1974})}\BibitemShut {NoStop}%
\end{thebibliography}%

\end{document}